\documentclass[aip,jcp,preprint,superscriptaddress,nofootinbib]{revtex4-1}
\usepackage{silence}
\usepackage[english]{babel}
\usepackage[T1]{fontenc}
\usepackage[utf8]{inputenc}
\usepackage{microtype}

\usepackage{amsmath}
\usepackage{amssymb}
\usepackage{textcomp}
\usepackage{textgreek}
\usepackage{physics}
\usepackage{upgreek}
\usepackage{mathtools}
\usepackage{bbm}
\usepackage{bbold}
\usepackage{bm}
\usepackage{mathrsfs}
\usepackage{nicefrac}

\usepackage{graphicx}
\usepackage{xcolor}
\usepackage{float}
\usepackage{booktabs}
\usepackage{enumitem}
\usepackage[colorlinks]{hyperref}
\usepackage{braket}
\usepackage{array}
\usepackage{siunitx}
\usepackage{threeparttable}
\usepackage{lipsum}

\newcommand{\mrm}[1]{\mathrm{#1}}
\newcommand{\mbf}[1]{\mathbf{#1}}
\newcommand{\nn}{\nonumber\\}
\newcommand{\trans}{\mspace{-3mu}\mathsf{T}}

\newcommand{\bigbraket}[2]{\Big\langle #1 \mspace{2mu} \Big\vert \mspace{1.5mu} #2 \Big\rangle}

\newcommand{\2}{\mspace{-2mu}}
\newcommand{\elm}[3]{\langle #1 \vert #2\vert #3 \rangle}

\makeatletter
\renewenvironment{dcases}[1][l]{\matrix@check\dcases\env@dcases{#1}}{\endarray\right.}
\def\env@dcases#1{%
  \let\@ifnextchar\new@ifnextchar
  \left\lbrace\def\arraystretch{1.2}%
  \array{@{}#1@{\quad}l@{}}}
\makeatother

\newcommand{\Eplain}[3]{E^{#1}_{{#2}^{#1} {#3}^{#1}}}

\newcommand{\Eplainprime}[3]{E^{#1'}_{{#2}^{#1^{\smash{\prime}}} \! {#3}^{#1^{\smash{\prime}}}}}

\newcommand{\Dplain}[3]{D^{#1}_{{#2}^{#1} {#3}^{#1}}}

\newcommand{\Dplainprime}[3]{D^{#1'}_{{#2}^{#1^{\smash{\prime}}} \! {#3}^{#1^{\smash{\prime}}}}}

\newcommand{\kappaplain}[3]{\kappa^{#1}_{{#2}^{#1} {#3}^{#1}}}

\newcommand{\gplain}[3]{g^{#1}_{{#2}^{#1} {#3}^{#1}}}

\newcommand{\fplain}[3]{f_{(#1 \, {#2}^{#1} {#3}^{#1})}}
\newcommand{\ftildeplain}[3]{\tilde{f}_{(#1 \, {#2}^{#1} {#3}^{#1})}}

\newcommand{\dplain}[5]{d^{#1}_{({#2}^{#1} {#3}^{#1}) ({#4}^{#1} {#5}^{#1})}}

\newcommand{\dfullplain}[6]{d_{ ({#1} \, {#2}^{#1}  {#3}^{#1}) ({#4'} {#5}^{#4^{\smash{\prime}}} \! {#6}^{#4^{\smash{\prime}}}) } }

\newcommand{\Cplain}[6]{C_{ ({#1} \, {#2}^{#1}  {#3}^{#1}) ({#4'} {#5}^{#4^{\smash{\prime}}} \! {#6}^{#4^{\smash{\prime}}}) } }
\newcommand{\Ctildeplain}[6]{\tilde{C}_{ ({#1} \, {#2}^{#1}  {#3}^{#1}) ({#4'} {#5}^{#4^{\smash{\prime}}} \! {#6}^{#4^{\smash{\prime}}}) } }

\newcommand{\Aprime}[4]{A_{ {#1} ({#2'} {#3}^{#2^{\smash{\prime}}} \! {#4}^{#2^{\smash{\prime}}}) } }

\newcommand{\Atildeprime}[4]{\tilde{A}_{ {#1} ({#2'} {#3}^{#2^{\smash{\prime}}} \! {#4}^{#2^{\smash{\prime}}}) } }

\newcommand{\aplain}[2]{a^{#1\mspace{-0.5mu}\raisebox{0.3ex}{$\scriptstyle\dagger$}}_{\mspace{-0mu}#2^{#1}}}

\newcommand{\bplain}[2]{b^{#1}_{\mspace{-1mu}#2^{#1}}}

 \usepackage{acro}

\DeclareAcronym{mctdh}{
   short = MCTDH ,
   long = multiconfiguration time-dependent Hartree ,
}

\DeclareAcronym{nomctdh}{
   short = NOMCTDH ,
   long = non-orthogonal \ac{mctdh} ,
}

\DeclareAcronym{gmctdh}{
   short = G-MCTDH ,
   long = Gaussian-based \ac{mctdh} ,
}

\DeclareAcronym{mlgmctdh}{
   short = ML-GMCTDH ,
   long = multilayer Gaussian-based \ac{mctdh} ,
}

\DeclareAcronym{mlmctdh}{
   short = ML-MCTDH ,
   long = multilayer \ac{mctdh} ,
}

\DeclareAcronym{mpsmctdh}{
   short = MPS-MCTDH ,
   long = matrix product state \ac{mctdh} ,
}

\DeclareAcronym{vmcg}{
   short = vMCG ,
   long = variational multiconfiguration Gaussian ,
}

\DeclareAcronym{ms}{
   short = MS ,
   long = multiple spawning ,
}

\DeclareAcronym{ccs}{
   short = CCS ,
   long = coupled coherent states ,
}

\DeclareAcronym{mctdhn}{
   short = MCTDH[\textit{n}] ,
   long = systematically truncated multiconfiguration time-dependent Hartree ,
}

\DeclareAcronym{mrmctdhn}{
  short = MR-MCTDH[\textit{n}] ,
  long = multi-reference truncated multiconfiguration time-dependent Hartree ,
}

\DeclareAcronym{tdh}{
   short = TDH ,
   long = time-dependent Hartree ,
}

\DeclareAcronym{xtdh}{
   short = X-TDH ,
   long = exponentially parametrized time-dependent Hartree ,
}

\DeclareAcronym{dmrg}{
   short = DMRG ,
   long = density matrix renormalization group,
}

\DeclareAcronym{tddmrg}{
   short = TD-DMRG ,
   long = time-dependent density matrix renormalization group,
}

\DeclareAcronym{scf}{
   short = SCF ,
   long = self-consistent field ,
}

\DeclareAcronym{casscf}{
   short = CASSCF ,
   long = complete active space self-consistent field ,
}

\DeclareAcronym{tdcasscf}{
   short = TD-CASSCF ,
   long = time-dependent \acl{casscf} ,
}

\DeclareAcronym{gasscf}{
   short = CASSCF ,
   long = generalized active space self-consistent field ,
}

\DeclareAcronym{tdgasscf}{
   short = TD-GASSCF ,
   long = time-dependent \acl{gasscf} ,
}

\DeclareAcronym{rasscf}{
   short = RASSCF ,
   long = restricted active space self-consistent field ,
}

\DeclareAcronym{tdrasscf}{
   short = TD-RASSCF ,
   long = time-dependent \acl{rasscf} ,
}

\DeclareAcronym{ormas}{
   short = ORMAS ,
   long = occupation-restricted multiple active space ,
}

\DeclareAcronym{tdormas}{
   short = TD-ORMAS ,
   long = time-dependent \acl{ormas} ,
}

\DeclareAcronym{mctdhf}{
   short = MCTDHF ,
   long = multiconfiguration time-dependent Hartree-Fock ,
}

\DeclareAcronym{tdhf}{
   short = TDHF ,
   long = time-dependent Hartree-Fock ,
}

\DeclareAcronym{occ}{
   short = OCC ,
   long = orbital-optimized coupled cluster ,
}

\DeclareAcronym{tdocc}{
   short = TD-OCC ,
   long = time-dependent \acl{occ} ,
}

\DeclareAcronym{nocc}{
   short = NOCC ,
   long = non-orthogonal orbital-optimized coupled cluster ,
}

\DeclareAcronym{oatdcc}{
   short = OATDCC ,
   long = orbital-adaptive time-dependent coupled cluster ,
}

\DeclareAcronym{fci}{
   short = FCI ,
   long = full configuration interaction ,
}

\DeclareAcronym{cud}{
   short = CUD ,
   long = closed under de-exciation ,
}

\DeclareAcronym{fsmr}{
   short = FSMR ,
   long = full-space matrix representation ,
}

\DeclareAcronym{hh}{
   short = HH ,
   long = H\'enon-Heiles ,
}

\DeclareAcronym{ho}{
   short = HO ,
   long = harmonic oscillator ,
}

\DeclareAcronym{dop853}{
   short = DOP853 ,
   long = Dormand-Prince 8{(5,3)} ,
}

\DeclareAcronym{sm}{
   short = SM ,
   long = supplementary material ,
}

\DeclareAcronym{vscf}{
   short = VSCF ,
   long = vibrational self-consistent field ,
}

\DeclareAcronym{eom}{
   short = EOM ,
   long = equation of motion ,
   short-plural-form = EOMs ,
   long-plural-form = equations of motion ,
   foreign-plural={}
}

\DeclareAcronym{tdvp}{
   short = TDVP ,
   long = time-dependent variational principle
}

\DeclareAcronym{tdse}{
   short = TDSE ,
   long = time-dependent Schr{\"o}dinger equation ,
}

\DeclareAcronym{cc}{
   short = CC ,
   long = coupled cluster ,
}

\DeclareAcronym{vcc}{
   short = VCC ,
   long = vibrational coupled cluster ,
}

\DeclareAcronym{tdvcc}{
   short = TDVCC ,
   long = time-dependent vibrational coupled cluster ,
}

\DeclareAcronym{tdvci}{
   short = TDVCI ,
   long = time-dependent vibrational configuration interaction ,
}

\DeclareAcronym{vci}{
   short = VCI ,
   long = vibrational configuration interaction ,
}

\DeclareAcronym{ci}{
   short = CI ,
   long = configuration interaction ,
}

\DeclareAcronym{tdci}{
   short = CI ,
   long = time-dependent \acl{ci} ,
}

\DeclareAcronym{sq}{
   short = SQ ,
   long = second quantization ,
}

\DeclareAcronym{fq}{
   short = FQ ,
   long = first quantization ,
}

\DeclareAcronym{mc}{
   short = MC ,
   long = mode combination ,
}

\DeclareAcronym{mcr}{
   short = MCR ,
   long = mode combination range ,
   long-plural = s ,
}

\DeclareAcronym{pes}{
   short = PES ,
   long = potential energy surface
}

\DeclareAcronym{svd}{
   short = SVD ,
   long = singular value decomposition ,
}

\DeclareAcronym{adga}{
   short = ADGA ,
   long = adaptive density-guided approach ,
}

\DeclareAcronym{rhs}{
   short = RHS ,
   long = right-hand side ,
}

\DeclareAcronym{lhs}{
   short = LHS ,
   long = left-hand side ,
}

\DeclareAcronym{ivr}{
   short = IVR ,
   long = intramolecular vibrational energy redistribution ,
}

\DeclareAcronym{fft}{
   short = FFT ,
   long = fast Fourier transform ,
}

\DeclareAcronym{spf}{
   short = SPF ,
   long = single-particle function ,
}

\DeclareAcronym{lls}{
   short = LLS ,
   long = linear least squares ,
}

\DeclareAcronym{itnamo}{
   short = ItNaMo ,
   long = iterative natural modal ,
}

\DeclareAcronym{hf}{
   short = HF ,
   long = Hartree-Fock ,
}

\DeclareAcronym{mcscf}{
   short = MCSCF ,
   long = multi-configurational self-consistent field ,
}

\DeclareAcronym{sop}{
   short = SOP ,
   long = sum-of-products ,
}

\DeclareAcronym{midascpp}{
   short = MidasCpp ,
   long = Molecular Interactions{,} Dynamics And Simulations Chemistry Program Package ,
}

\DeclareAcronym{mpi}{
   short = MPI ,
   long = message passing interface ,
}

\DeclareAcronym{ode}{
   short = ODE ,
   long  = ordinary differential equation ,
   short-plural = s ,
   long-plural = s ,
   short-indefinite = an ,
   long-indefinite = an ,
   foreign-plural={}
}

\DeclareAcronym{bch}{
   short = BCH ,
   long = Baker-Campbell-Hausdorff ,
}

\DeclareAcronym{sr}{
   short = SR ,
   long = single-reference ,
}

\DeclareAcronym{mr}{
   short = MR ,
   long = multi-reference ,
}

\DeclareAcronym{dof}{
   short = DOF ,
   long = degree of freedom ,
   short-plural-form = DOFs ,
   long-plural-form = degrees of freedom ,
}

\DeclareAcronym{hp}{
   short = HP ,
   long = Hartree product ,
}

\DeclareAcronym{tdbvp}{
   short = TDBVP ,
   long  = time-dependent bivariational principle ,
   short-plural = s ,
   long-plural = s ,
   short-indefinite = a ,
   long-indefinite = a ,
}

\DeclareAcronym{dfvp}{
   short = DFVP ,
   long  = Dirac-Frenkel variational principle ,
}

\DeclareAcronym{ele}{
   short = ELE ,
   long  = Euler-Lagrange equation ,
   short-plural = s ,
   long-plural = s ,
   foreign-plural={}
}

\DeclareAcronym{mrcc}{
   short = MRCC ,
   long = multi-reference coupled cluster ,
}

\DeclareAcronym{tdfvci}{
   short = TDFVCI ,
   long = time-dependent full vibrational configuration interaction ,
}

\DeclareAcronym{tdfci}{
   short = TDFCI ,
   long = time-dependent full configuration interaction ,
}

\DeclareAcronym{tdevcc}{
   short = TDEVCC ,
   long  = time-dependent extended vibrational coupled cluster ,
   short-plural = s ,
   long-plural = s ,
   short-indefinite = a ,
   long-indefinite = a ,
}

\DeclareAcronym{holc}{
   short = HOLC ,
   long = hybrid optimized and localized vibrational coordinate ,
}

\DeclareAcronym{acf}{
   short = ACF ,
   long = autocorrelation function ,
   foreign-plural={}
}

\DeclareAcronym{fwhm}{
   short = FWHM ,
   long  = full width at half maximum ,
   short-plural = s ,
   long-plural = full widths at half maxima ,
   short-indefinite = an ,
   long-indefinite = a ,
}

\DeclareAcronym{tdmvcc}{
   short = TDMVCC ,
   long = time-dependent vibrational coupled cluster with time-dependent modals ,
}

\DeclareAcronym{midas}{
   short = MidasCpp ,
   long = Molecular Interactions{,} Dynamics and Simulations Chemistry Program Package ,
} 

\usepackage[normalem]{ulem}
\newcommand{\stkout}[1]{\ifmmode\text{\sout{\ensuremath{#1}}}\else\sout{#1}\fi}
\newcommand{\revision}[3][]{#3}

\newcommand{\au}{Department of Chemistry, University of Aarhus, \\ Langelandsgade 140, DK--8000 Aarhus C, Denmark}
\newcommand{\upo}{Dipartimento di Scienze e Innovazione Tecnologica, Universit\`a del Piemonte Orientale (UPO), Via T. Michel 11, 15100 Alessandria, Italy}

\begin{document}

\title{General exponential basis set parametrization: Application to time-dependent bivariational wave functions}
\author{Mads Greisen Højlund}
\email{madsgh@chem.au.dk}
\affiliation{\au}
\author{Alberto Zoccante}
\email{alberto.zoccante@uniupo.it}
\affiliation{\upo}
\author{Ove Christiansen}
\email{ove@chem.au.dk}
\affiliation{\au}

\hypersetup{pdftitle={General exponential basis set parametrization: Application to time-dependent bivariational wave functions}}
\hypersetup{pdfauthor={M.~G.~Højlund, et al.}}
\hypersetup{bookmarksopen=true}

\date{\today}


\begin{abstract}
    We present \acp{eom} for general time-dependent wave functions
with exponentially parametrized biorthogonal basis sets. The equations
are fully bivariational in the sense of the \ac{tdbvp} and offer
an alternative, constraint free formulation of adaptive basis sets for
bivariational wave functions.
We simplify the highly non-linear basis set equations using
Lie algebraic techniques and show that the computationally 
intensive parts of the theory are in fact identical to those that arise
with linearly parametrized basis sets. Our approach thus offers easy implementation
on top of existing code
in the context
of both nuclear dynamics and time-dependent electronic structure.
Computationally tractable working equations are provided for single and 
double exponential parametrizations of the basis set evolution. 
The \acp{eom} are generally applicable for any value of the basis set
parameters, unlike the approach of transforming the parameters to zero at each evaluation of the \acp{eom}.
%
We show that the basis set equations contain a well-defined set
of singularities, which are identified and removed by a simple scheme.
The exponential basis set equations are implemented in conjunction
with \ac{tdmvcc} and we investigate the propagation properties
in terms of the average integrator step size. For the systems we test,
the exponentially parametrized basis sets yield slightly larger
step sizes compared to linearly parametrized basis set. 

 \end{abstract}

\maketitle

\acresetall

\section{Introduction} \label{sec:introduction}
Optimized basis sets are ubiquitous in time-independent quantum chemistry, 
since the proper choice of basis is crucial to a compact representation of the wave function.
The paradigmatic example in electronic structure theory is the \ac{scf} method, while
vibrational \ac{scf} (\acs{vscf}) plays a similar role in vibrational structure theory.
Correlated electronic wave function methods that simultaneously optimize the basis set are also
in widespread use, with multi-configurational \ac{scf} (\acs{mcscf}) as the standard example
and
\ac{dmrg} with self-consistently optimized orbitals
(DMRG-SCF)\cite{zgidDensityMatrixRenormalization2008,ghoshOrbitalOptimizationDensity2008,maSecondOrderSelfConsistentFieldDensityMatrix2017}
as a more recent example.
Basis set optimization also plays a role in electronic
\acl{cc} theory through
orthogonal\cite{sherrillEnergiesAnalyticGradients1998,pedersenGaugeInvariantCoupled1999}
and non-orthogonal\cite{pedersenGaugeInvariantCoupled2001} orbital-optimized
coupled cluster (\acs{occ} and \acs{nocc}).
These and similar methods require special formal considerations
since they are not variational (in the sense of providing an upper bound on the energy) and
since the bra and ket states are parametrized in an asymmetric fashion.
A principled way of handling this asymmetry while ensuring convergence to \ac{fci}
is to use the bivariational framework pioneered
by Arponen\cite{arponenVariationalPrinciplesLinkedcluster1983}. The \ac{nocc} method is fully bivariational
and thus converges\cite{myhreDemonstratingThatNonorthogonal2018} to \ac{fci}, while
\ac{occ} does not\cite{kohnOrbitaloptimizedCoupledclusterTheory2005}.

Optimized or adaptive basis sets are also important in explicitly time-dependent theory as
illustrated by the success of
methods such as \ac{mctdh}\cite{meyerMulticonfigurationalTimedependentHartree1990,beckMulticonfigurationTimedependentHartree2000}
for nuclear dynamics and the related \ac{mctdhf}\cite{zanghelliniMCTDHFApproachMulti2003,caillatCorrelatedMultielectronSystems2005}
for electronic dynamics.
\revision[MGH]{}{%
Various schemes that use adaptive basis sets (like \ac{mctdh} and \ac{mctdhf}) 
but restrict or truncate the wave function expansion are known as well.
\cite{katoTimedependentMulticonfigurationTheory2004, 
satoTimedependentCompleteactivespaceSelfconsistentfield2013, 
miyagiTimedependentRestrictedactivespaceSelfconsistentfield2013, 
haxtonTwoMethodsRestricted2015a, 
satoTimedependentMulticonfigurationSelfconsistentfield2015, 
worthAccurateWavePacket2000, 
wodraszkaUsingPrunedNondirect2016, 
larssonDynamicalPruningMulticonfiguration2017, 
madsenSystematicVariationalTruncation2020, 
madsenMRMCTDHFlexible2020} 
Methods that treat mixed particle types (e.g. fermionic and bosonic) also exist.
\cite{alonMulticonfigurationalTimedependentHartree2007,
alonMulticonfigurationalTimedependentHartree2008,
nestMulticonfigurationElectronNuclear2009,
mantheMultilayerMulticonfigurationalTimedependent2017}
}

There has been considerable interest in formulating theories that combine the idea
of adaptive basis sets with the use of asymmetric parametrizations such as coupled cluster.
Although such theories can be quite involved, several works have appeared in
the context of electronic\cite{kvaalInitioQuantumDynamics2012,satoCommunicationTimedependentOptimized2018,pedersenSymplecticIntegrationPhysical2019,kristiansenNumericalStabilityTimedependent2020,pathakTimedependentOptimizedCoupledcluster2020,pathakTimedependentOptimizedCoupledcluster2021,kristiansenLinearNonlinearOptical2022}
and nuclear\cite{madsenTimedependentVibrationalCoupled2020,hojlundBivariationalTimedependentWave2022} dynamics.
Some of these works (as well as the present work) utilize Arponen's \ac{tdbvp}\cite{arponenVariationalPrinciplesLinkedcluster1983},
which offers a clear formal strategy for deriving \acp{eom}.
However, additional work is typically needed to obtain the \acp{eom} in a computationally tractable format
and to analyze the \acp{eom} for redundancies and singularities.

It is well known that unitary (invertible) basis set transformations can be parametrized in terms of the exponential
of an anti-Hermitian (general) one-particle operator $\revision[MGH]{}{\hat{\kappa}}$ (see Ref. \citenum{helgakerMolecularElectronicstructureTheory2000} for an overview of
the standard mathematical machinery). This constraint free approach results in simple equations when expanding around $\revision[MGH]{}{\hat{\kappa}} = 0$
and has been used frequently in the context of time-independent theory and response theory where one can trivially
obtain $\revision[MGH]{}{\hat{\kappa}} = 0$ by absorbing a non-zero $\revision[MGH]{}{\hat{\kappa}}$ into the Hamiltonian integrals. 
In an explicitly time-dependent context the situation is more complicated and a linear
parametrization with constraints has usually been preferred over the exponential parametrization.
However, a few exceptions can be found in the literature:
Pedersen and Koch\cite{pedersenTimedependentLagrangianApproach1998}
presented general considerations on exponentially parametrized \ac{tdhf}
but did not derive explicit \acp{eom} for $\revision[MGH]{}{\hat{\kappa}}$.
Madsen et al.\cite{madsenExponentialParameterizationWave2018} derived and implemented the
\acp{eom} for \ac{xtdh} without assuming $\revision[MGH]{}{\hat{\kappa}} = 0$. In this case the exponential
parametrization resulted in substantial computational gains compared to the conventional
linear parametrization. 
Recently, Kristiansen et al.\cite{kristiansenLinearNonlinearOptical2022}
considered time-dependent \ac{occ} and \ac{nocc} with double excitations (TDOCCD and TDNOCCD)
and introduced the corresponding second-order approximations
TDOMP2 and TDNOMP2 with exponentially parametrized orbitals.
They used the $\revision[MGH]{}{\hat{\kappa}} = 0$ method to simplify derivations
while keeping the benefit of a contraint free formulation. 
\revision[MGH]{}{%
We note that the exponential parametrization with $\revision[MGH]{}{\hat{\kappa}} = 0$ and the linear parametrization with constraints
result in very similar working equations although the derivations have different
starting points. Allowing $\revision[MGH]{}{\hat{\kappa}} \neq 0$ leads to substantially different equations.
}

The question of how to parametrize basis sets is not only one of mathematical convenience---
it also plays an important role in the propagation of time-dependent
wave functions. This inludes questions of numerical stability and integrator step size, the latter being
a determining factor of the computational cost. 

In a recent paper\cite{hojlundBivariationalTimedependentWave2022}
we considered linear basis set parametrization and a novel parametrization based on polar
decomposition for general time-dependent bivariational wave functions.
The purpose of that work was mainly to study the effect on the numerical
stability of \ac{tdmvcc}\cite{madsenTimedependentVibrationalCoupled2020}
and it was shown that a so-called restricted polar parametrization
offered improved numerical stability compared to the linear parametrization.
The present work considers the \acp{eom} for similar kinds of wave functions
with exponentially parametrized basis sets without assuming $\revision[MGH]{}{\hat{\kappa}} = 0$.
In addition to the single exponential formalism, we also
consider a double exponential basis set parametrization. The former
corresponds to the linear parametrization, while the latter parallels the polar
parametrization of Ref. \citenum{hojlundBivariationalTimedependentWave2022}.
The \acp{eom} derived are general with respect to the type of wave function expansion but
the main application is \acl{cc} theory where a bivariational formulation is natural.
We note that the derivations are mainly presented in the language and notation
of vibrational structure theory but that all theoretical
results carry over
to electronic structure theory after minor notational adjustments such as dropping
mode indices and sums over modes. 

The paper is organized as follows: Section~\ref{sec:theory}
covers the theory, including a brief introduction to the
\ac{tdbvp} and derivations of \acp{eom}. This is followed
by a description of our implementation 
for the nuclear dynamics case
in Sec.~\ref{sec:implementation}
and a few numerical examples in Sec.~\ref{sec:results}. 
Section \ref{sec:summary} concludes the paper with a summary of our findings and
an outlook on future work. 

\section{Theory} \label{sec:theory}

\subsection{The time-dependent bivariational principle} \label{ssec:tdbvp}
Following Arponen\cite{arponenVariationalPrinciplesLinkedcluster1983}, we consider a
general bivariational Lagrangian,
\begin{align}
    \mathcal{L} = \elm{\Psi'}{(i \partial_t - H)}{\Psi},
\end{align}
where the bra and ket states are independent. We then determine stationary points ($\delta \mathcal{S} = 0$) 
of the action-like functional
\begin{align} \label{eq:tdbvp}
    \mathcal{S} = \int_{t_0}^{t_1} \mathcal{L} \dd{t}
\end{align}
under the condition that the variations of the bra and ket vanish at the endpoints of the integral.
In the exact-theory case, a short calculation shows that this procedure is equivalent to the \ac{tdse} and its complex conjugate.
For approximate, parametrized wave functions, the stationarity condition is instead equivalent to a set
of \acp{ele},
\begin{align} \label{eq:eles}
    0 = \pdv{\mathcal{L}}{y_i} - \dv{t} \pdv{\mathcal{L}}{\dot{y}_i},
\end{align}
for all parameters $y_i$ (a short proof of this well-known fact is given in Ref.~\citenum{hojlundBivariationalTimedependentWave2022} in a notation consistent with the present work). Writing the Lagrangian as
\begin{align}
    \mathcal{L} = \mathcal{I} - \mathcal{H}
\end{align}
with
\begin{subequations}
\begin{gather}
    \mathcal{I}(\mbf{y}, \dot{\mbf{y}}) = i \braket{\Psi' | \dot{\Psi}} = i \sum_{j} \dot{y}_j \bigbraket{\Psi' }{ \pdv{\Psi}{y}}, \\
    \mathcal{H}(\mbf{y}, t) = \elm{\Psi'}{H}{\Psi},
\end{gather}
\end{subequations}
the \acp{ele} in Eq.~\eqref{eq:eles} become
\begin{align}
    \pdv{\mathcal{I}}{y_i} - \dv{t} \pdv{\mathcal{I}}{\dot{y}_i} = \pdv{\mathcal{H}}{y_i}.
\end{align}
After carrying out the derivatives using the chain and product rules and cancelling terms,
one gets the following appealing \acp{eom}:
\begin{align} \label{eq:eom_simple}
    i \sum_{j} \mathcal{M}_{ij} \dot{y}_j &= \pdv{\mathcal{H}}{y_i}.
\end{align}
Here, we have defined an anti-symmetric matrix $\bm{\mathcal{M}}$ 
with elements
\begin{equation} \label{eq:full_M_matrix_def}
    \mathcal{M}_{ij} = 
    \bigbraket{\pdv{\Psi'}{y_i}}{\pdv{\Psi}{y_j}} -
    \bigbraket{\pdv{\Psi'}{y_j}}{\pdv{\Psi}{y_i}}. 
\end{equation}
The \acp{eom} in Eq.~\eqref{eq:eom_simple} consititute a natural bivariational analogue
of the general variational \acp{eom} considered by e.g. Kramer and Saraceno\cite{kramerGeometryTimeDependentVariational1981}
and Ohta\cite{ohtaTimedependentVariationalPrinciple2004}.
 
\subsection{Parametrization} \label{ssec:parametrization}
We will consider general wave function parametrizations of the form
\begin{subequations}
\begin{align}
    \ket{\Psi  (\bm{\alpha}, \bm{\kappa})}  &= e^{\revision[MGH]{}{\hat{\kappa}}} \ket{\psi (\bm{\alpha})} \\
    \bra{\Psi' (\bm{\alpha}, \bm{\kappa})} &= \bra{\psi' (\bm{\alpha})} e^{-\revision[MGH]{}{\hat{\kappa}}}
\end{align}
\end{subequations}
where the one-particle operator $\revision[MGH]{}{\hat{\kappa}}$ generates invertible linear transformations 
of the \revision[MGH]{}{basis functions} (modals or orbitals). In the vibrational case, we 
use the notation of many-mode second quantization\cite{christiansenSecondQuantizationFormulation2004} and write
\begin{align}
    \revision[MGH]{}{\hat{\kappa}} = \sum_{m} \sum_{p^m q^m} \kappaplain{m}{p}{q} \Eplain{m}{p}{q}.
\end{align}
The parameters $\kappaplain{m}{p}{q}$ are collected
in vectors $\bm{\kappa}^m$ or matrices $\mbf{K}^m$
depending on context.
The one-mode shift operators,
\begin{align}
    \Eplain{m}{p}{q} = \aplain{m}{p} \bplain{m}{q},
\end{align}
satisfy the commutator
\begin{align} \label{eq:E_commutator}
    [\Eplain{m}{p}{q}, \Eplainprime{m}{r}{s}] = \delta_{mm'} (\delta_{q^m r^m} \Eplain{m}{p}{s} - \delta_{p^m s^m} \Eplain{m}{r}{q})
\end{align}
and constitute the generators of the general linear group. 
The well-known one-electron shift operators 
satisfy essentially the same commutator\cite{helgakerMolecularElectronicstructureTheory2000}, which means that the derivations in this work carry over to electronic
structure theory after removing mode indices as appropriate.
We will denote the remaining wave function parameters $\bm{\alpha}$ as configurational
parameters. The non-transformed states $\ket{\psi (\bm{\alpha})}$
and $\bra{\psi' (\bm{\alpha})}$ are assumed to be expressed in the same primitive basis as the $\revision[MGH]{}{\hat{\kappa}}$ operator.
We will allow for the case where the wave function is expanded in an active subset of the primitive basis.
In Sec.~\ref{ssec:eom} we simply require that the primitive basis is biorthonormal, while we require
orthonormality in Sec.~\ref{ssec:double_exp}. We note that this
restriction can likely be lifted with appropriate modifications of the derivations.
 
\subsection{Equations of motion} \label{ssec:eom}
Ordering the parameters like $\mbf{y} = (\bm{\alpha}, \bm{\kappa})$, the \acp{eom} in Eq.~\eqref{eq:eom_simple}
assume the structure
\begin{align} \label{eq:eom_block}
    i
    \left[
    \begin{array}{c | c} 
    \mbf{M}              & {\mspace{8mu}\tilde{\mbf{A}}\mspace{5mu}} 
    \\ \hline
    -\tilde{\mbf{A}}^{\2\trans}  & {\mspace{8mu}\tilde{\mbf{C}}\mspace{5mu}}  
    \end{array} \right]
    \left[
    \begin{array}{c} 
    \dot{\bm{\alpha}}
    \\ \hline
    \dot{\bm{\kappa}}
    \end{array} \right]
    =
    \left[
    \begin{array}{c} 
    \mbf{h}
    \\ \hline
    \tilde{\mbf{f}}
    \end{array} \right].
\end{align}
The various matrix elements can be calculated by direct application of Eq.~\eqref{eq:full_M_matrix_def}
followed by the appropriate use of
\begin{subequations}
\begin{align}
    e^{\revision[MGH]{}{\hat{\kappa}}} e^{-\revision[MGH]{}{\hat{\kappa}}} &= 1, \\
    \pdv{e^{-\revision[MGH]{}{\hat{\kappa}}}}{\kappaplain{m}{p}{q}} e^{\revision[MGH]{}{\hat{\kappa}}} &= - e^{-\revision[MGH]{}{\hat{\kappa}}} \pdv{e^{\revision[MGH]{}{\hat{\kappa}}}}{\kappaplain{m}{p}{q}}
\end{align}
\end{subequations}
where the latter follows from taking the derivative of the former. For notational convenience, we
define a similarity transformed derivative\cite{hallLieGroupsLie2015,olsenLinearNonlinearResponse1985}
\begin{align} \label{eq:d_oper_def}
	\Dplain{m}{p}{q} 
    &= e^{-\revision[MGH]{}{\hat{\kappa}}} \pdv{e^{\revision[MGH]{}{\hat{\kappa}}}}{\kappaplain{m}{p}{q}} \nn
    &= e^{-\revision[MGH]{}{\hat{\kappa}}^m} \pdv{e^{\revision[MGH]{}{\hat{\kappa}}^m}}{\kappaplain{m}{p}{q}} \nn
	&= \sum_{n=0}^{\infty} \frac{(-1)^n}{(n+1)!} (\mrm{ad}_{\revision[MGH]{}{\hat{\kappa}}^m})^n \Eplain{m}{p}{q} \nn
	&= \sum_{r^m s^m} \dplain{m}{p}{q}{r}{s} \Eplain{m}{r}{s}
\end{align}
with $\mrm{ad}_{\revision[MGH]{}{\hat{\kappa}}^m} (X) = [\revision[MGH]{}{\hat{\kappa}}^m, X]$
(the second equality follows from the fact that $[\revision[MGH]{}{\hat{\kappa}}^m, \revision[MGH]{}{\hat{\kappa}}^{m'}] = 0$ for $m \neq m'$).
This operator is a one-particle operator as indicated by the
last equality in Eq.~\eqref{eq:d_oper_def}. For now we assume that the matrix elements $\dplain{m}{p}{q}{r}{s}$
are available and collect them in a matrix $\mbf{D}^m$. We will later show how the elements of $\mbf{D}^m$ can be calculated
and, in particular, how transformations by $\mbf{D}^m$ can be performed in an efficient manner
without explicitly constructing the full matrix.
The matrix elements needed for Eq.~\eqref{eq:eom_block} are now given by
\begin{subequations}
    \begin{align} \label{eq:M_A_C_h_f_def_with_tilde}
        M_{ij} &= 
        \bigbraket{\pdv{\psi'}{\alpha_i}}{\pdv{\psi}{\alpha_j}} -
        \bigbraket{\pdv{\psi'}{\alpha_j}}{\pdv{\psi}{\alpha_i}} \\
        \Atildeprime{i}{m}{r}{s} &= \pdv{}{\alpha_i} \elm{\psi'}{\Dplainprime{m}{r}{s}}{\psi} \\
        \Ctildeplain{m}{p}{q}{m}{r}{s} &= \elm{\psi'}{[\Dplainprime{m}{r}{s}, \Dplain{m}{p}{q}]}{\psi} \\
        h_i &= 
        \pdv{}{\alpha_i} \elm{\psi'}{\bar{H}}{\psi} \label{eq:h_element} \\
        \ftildeplain{m}{p}{q} &= 
        \elm{\psi'}{[\bar{H}, \Dplain{m}{p}{q}]}{\psi}
    \end{align}
\end{subequations}
where we have defined
\begin{equation}
    \bar{H} = e^{-\revision[MGH]{}{\hat{\kappa}}} H e^{\revision[MGH]{}{\hat{\kappa}}}
\end{equation}
and used a tilde to indicate the quantities that depend on the $\Dplain{m}{p}{q}$
operators.
These quantities can be quite complicated since $\Dplain{m}{p}{q}$ is generally a
full one-particle operator. In particular, one cannot immediately analyze the elements of 
$\tilde{\mbf{A}}$ and $\tilde{\mbf{C}}$ for zeros.
Having calculated the matrix elements of the \acp{eom}, it is easy 
to show that Eq.~\eqref{eq:eom_block} is equivalent to
\begin{subequations} \label{eq:eom_inverted}
    \begin{align}
        i \dot{\bm{\alpha}} &= \mbf{M}^{-1} \big(\mbf{h} - i \tilde{\mbf{A}} \dot{\bm{\kappa}} \big), \label{eq:eom_alpha} \\
        i \big(\tilde{\mbf{C}} + \tilde{\mbf{A}}^{\trans} \mbf{M}^{-1} \tilde{\mbf{A}} \big) \dot{\bm{\kappa}}
        &= \tilde{\mbf{f}} + \tilde{\mbf{A}}^{\trans} \mbf{M}^{-1} \mbf{h}. \label{eq:eom_kappa_dot}
    \end{align}
\end{subequations}
if $\mbf{M}$ is invertible.
\revision[MGH]{}{%
The $\mathbf{M}$ matrix is usually rather trivial in a way
that does not depend on the concrete parameter values but rather on
the type of parametrization. 
Two examples (coupled cluster and linearly expanded wave functions)
are given in Ref.~\citenum{hojlundBivariationalTimedependentWave2022}. 
In those cases, the $\mathbf{M}$ matrix
and its inverse are simply given by
\begin{align} \label{eq:M_and_M_inv}
   \mathbf{M} = 
   \left[
  \begin{array}{c | c} 
  \mathbf{0}  & - \mathbf{1} 
  \\ \hline
  +\mathbf{1}  & \mathbf{0}  
  \end{array} \right]
   , \quad
   \mathbf{M}^{-1} = 
   \left[
  \begin{array}{c | c} 
  \mathbf{0}  & +\mathbf{1} 
  \\ \hline
  -\mathbf{1}  & \mathbf{0}  
  \end{array} \right].
\end{align}
For extended coupled cluster,
the $\mathbf{M}$ matrix is not trivial but it can still be inverted analytically
and without singularities.\cite{hansenExtendedVibrationalCoupled2020}
We are currently not aware of any wave function where $\mathbf{M}$ can become singular.}
\revision[MGH]{}{Instead, the} main difficulty is that
Eq.~\eqref{eq:eom_kappa_dot} cannot generally (i.e. for $\revision[MGH]{}{\hat{\kappa}} \neq 0$) 
be solved as it stands due to the presence of redundancies that
do not simply consist of certain blocks being equal to zero.
We can circumvent this problem by introducing appropriate zeroth 
order quantities, i.e.
\begin{subequations} \label{eq:A_C_f_def_without_tilde}
    \begin{align} 
        \Aprime{i}{m}{r}{s} &= \pdv{}{\alpha_i} \elm{\psi'}{\Eplainprime{m}{r}{s}}{\psi}, \label{eq:A_simple_def} \\
        \Cplain{m}{p}{q}{m}{r}{s} &= \elm{\psi'}{[\Eplainprime{m}{r}{s}, \Eplain{m}{p}{q}]}{\psi}, \\
        \fplain{m}{p}{q} &= 
        \elm{\psi'}{[\bar{H}, \Eplain{m}{p}{q}]}{\psi}.
    \end{align}
\end{subequations}
These are simpler to calculate and, more importantly, allow
a straight-forward (although possibly tedious) identification of
vanishing matrix elements through analysis of the shift operator expressions.
The introduction of a block-diagonal matrix $\mbf{D}$ with elements
\begin{align}
    \dfullplain{m}{p}{q}{m}{r}{s} = \delta_{mm'} \dplain{m}{p}{q}{r}{s}
\end{align}
now allows us to write
\begin{subequations} \label{eq:A_C_f_tilde_from_plain}
    \begin{align}
        \tilde{\mbf{A}} &= \mbf{A} \mbf{D}^{\trans}, \\
        \tilde{\mbf{C}} &= \mbf{D} \mbf{C} \mbf{D}^{\trans}, \\
        \tilde{\mbf{f}} &= \mbf{D} \mbf{f},
    \end{align}
\end{subequations}
thus relating the simple zeroth order matrices to the infinite order matrices.
Substituting Eqs.~\eqref{eq:A_C_f_tilde_from_plain} into Eqs.~\eqref{eq:eom_inverted}
then yields
\begin{subequations} \label{eq:eom_inverted_with_plain_matrices}
    \begin{align}
        i \dot{\bm{\alpha}} &= \mbf{M}^{-1} \big(\mbf{h} 
        - i \mbf{A} \mbf{D}^{\trans} \dot{\bm{\kappa}} \big), \label{eq:eom_alpha_with_plain_matrices} \\
        i \mbf{D} \big(\mbf{C}
        +   \mbf{A}^{\trans} \mbf{M}^{-1} \mbf{A} \big) \mbf{D}^{\trans} \dot{\bm{\kappa}}
        &= \mbf{D}
        \big( \mbf{f} + \mbf{A}^{\trans} \mbf{M}^{-1} \mbf{h} \big). \label{eq:eom_kappa_dot_with_plain_matrices}
    \end{align}
\end{subequations}
These equations are further simplified by removing $\mbf{D}$ on both sides
of Eq.~\eqref{eq:eom_kappa_dot_with_plain_matrices} and introducing the definition
\begin{align} \label{eq:kappa_tilde_dot_def}
    \mbf{g} = i \mbf{D}^{\trans} \dot{\bm{\kappa}}
\end{align} 
in order to write
\begin{subequations} \label{eq:eom_inverted_kappa_tilde_dot}
    \begin{align}
        i \dot{\bm{\alpha}} &= \mbf{M}^{-1} \big(\mbf{h} 
        - \mbf{A} \mbf{g} \big), \label{eq:eom_alpha_with_kappa_tilde_dot} \\
        \big(\mbf{C}
        +   \mbf{A}^{\trans} \mbf{M}^{-1} \mbf{A} \big) \mbf{g}
        &= 
        \big( \mbf{f} + \mbf{A}^{\trans} \mbf{M}^{-1} \mbf{h} \big). \label{eq:eom_kappa_tilde_dot}
    \end{align}
\end{subequations}
Equations~\eqref{eq:eom_inverted_kappa_tilde_dot} are the central equations
of the present work and determine the time evolution of the parameters.
Equation~\eqref{eq:eom_alpha_with_kappa_tilde_dot} directly gives the time derivative
of the configurational parameters, while Eq.~\eqref{eq:eom_kappa_tilde_dot}
determines the evolution of the basis set parameters in a slightly indirect way:
First the equation is solved, and then $\dot{\bm{\kappa}}$ is recovered from Eq.~\eqref{eq:kappa_tilde_dot_def}.
In the general case, this involves the inversion of $\mbf{D}^{\trans}$, while the computation
reduces to $\dot{\bm{\kappa}} = - i\mbf{g}$ in the $\revision[MGH]{}{\hat{\kappa}} = 0$ case.
It is interesting to note that Eqs.~\eqref{eq:eom_inverted_kappa_tilde_dot} are in fact
\textit{identical} to the central equations
for bivariational wave functions with linearly parametrized basis
sets [Eqs. (46) and (63) in Ref.~\citenum{hojlundBivariationalTimedependentWave2022}].
It that work, Eq.~\eqref{eq:eom_kappa_tilde_dot} appeared as a consequence of the 
necessary biorthonormality constraints and it was shown how it can be analyzed for redundancies
and solved for relevant types of wave functions, e.g. \acl{cc}.
The important point here is that the central and computationally intensive equations
are \textit{independent} of the choice of basis set parametrization.

As a final simplification we define
\begin{align}
    \mbf{h}' &= \mbf{h} - \mbf{A} \mbf{g}
\end{align}
and write the configurational \acp{eom} compactly as
\begin{align}
    i \dot{\bm{\alpha}} &= \mbf{M}^{-1} \mbf{h}'. \label{eq:eom_alpha_compact}
\end{align}
According to Eqs.~\eqref{eq:h_element} and \eqref{eq:A_simple_def} one may
calculate the elements of $\mbf{h}'$ as
\begin{align}
    h_i' = \pdv{}{\alpha_i} \elm{\psi'}{ ( \bar{H} - \revision[MGH]{}{\hat{g}} ) }{\psi} 
    = \pdv{\mathcal{H}'}{\alpha_i} 
\end{align}
with the one-particle operator $\revision[MGH]{}{\hat{g}}$
and the modified Hamiltonian function $\mathcal{H}'$ given by
\begin{align}
    \revision[MGH]{}{\hat{g}} &= \sum_{m} \sum_{p^m q^m} \gplain{m}{p}{q} \Eplain{m}{p}{q}, \\
    \mathcal{H}\revision[MGH]{}{'} &= \elm{\psi'}{ ( \bar{H} - \revision[MGH]{}{\hat{g}} ) }{\psi}.
\end{align}
Assuming that we are able to solve Eq.~\eqref{eq:eom_kappa_tilde_dot}
for $\mbf{g}$, we need to recover $\dot{\bm{\kappa}}$ in order
to integrate the \acp{eom}. Using the definition in Eq.~\eqref{eq:kappa_tilde_dot_def} 
and taking $\mbf{D}^{\trans}$ to be invertible, this means that we should compute
\begin{align}
    \dot{\bm{\kappa}}^m = -i (\mbf{D}^m)^{-\trans} \mbf{g}^m
\end{align}
for each mode. 
Doing this efficiently requires some general considerations on similarity
transformed derivatives such as the one in Eq.~\eqref{eq:d_oper_def}. These
considerations are covered by the next section.

\subsection{Similarity transformed derivative} \label{ssec:similarity_transformed_derivative}

In this section, we leave the notation of the main text and consider a general 
Lie group. We note that the following derivations are rather generic in a Lie group
context but we give the details for the convenience of the reader.
Somewhat similar considerations (although considerably less general) can be found
in Exercise 3.5 of Ref.~\citenum{helgakerMolecularElectronicstructureTheory2000}.

The generators of the group (i.e. the basis of the Lie algebra) will be denoted
by $E_i$ where $i$ might be a compound index. An element $X$ of the Lie algebra can then 
be written as
\begin{equation}
    X = X_i E_i
\end{equation}
where repeated indices imply summation (this convention is used throughout this section). 
The Lie algebra is characterized by the commutators (Lie brackets)
\begin{equation} \label{eq:lie_brackets}
    [E_i, E_j] = f^k_{ij} E_k
\end{equation}
where the scalars $f^k_{ij}$ are denoted structure constants. 
Equation~\eqref{eq:lie_brackets} is completely general and covers, e.g., the
one-particle shift operator commutator in Eq.~\eqref{eq:E_commutator}.
We will consider the case where
$X$ depends on a parameter $s$ through the coefficients $X_i$ and compute the quantity\cite{hallLieGroupsLie2015}
\begin{equation} \label{eq:main_general_d_oper_def}
    D = e^{-X(s)} \dv{s} e^{X(s)} = \sum_{k=0}^{\infty} \frac{(-1)^k}{(k+1)!} (\mathrm{ad}_X)^k \dot{X}
\end{equation}
where
\begin{gather}
    \mathrm{ad}_X (Y) = [X, Y], \\
    \dot{X} = \dv{X}{s} = \dv{X_i}{s} E_i = \dot{X}_i E_i.
\end{gather}
Before proceeding we note that
\begin{align} \label{eq:commutation_by_x}
    [X, E_i] = X_j [E_j, E_i] = X_j f^k_{ji} E_k = Q_{ik} E_k
\end{align}
where we have defined a matrix $\mbf{Q}$ with elements
\begin{equation}
    Q_{ik} = X_j f^{k}_{ji}.
\end{equation}
Equation~\eqref{eq:commutation_by_x} simply shows that commutation by $X$
translates to contraction with the matrix $\mbf{Q}$. Using this fact,
the first few commutators in the expansion become
\begin{subequations}
    \begin{align}
        (\mathrm{ad}_X)^0 \dot{X} 
        &= \dot{X}_i E_i \nn
        &= \dot{X}_i (\mbf{Q}^0)_{ij} E_j, \\
        (\mathrm{ad}_X)^1 \dot{X} 
        &= [X, \dot{X}_i E_i] \nn
        &= \dot{X}_i (\mbf{Q}^1)_{ij} E_j, \\
        (\mathrm{ad}_X)^2 \dot{X} 
        &= [X, [X, \dot{X}_i E_i]] \nn
        &= \dot{X}_i (\mbf{Q}^2)_{ij} E_j, \\
        &\vdotswithin{=} \nonumber
    \end{align}
\end{subequations}
Combining this 
pattern with Eq.~\eqref{eq:main_general_d_oper_def} yields an attractive expression, namely
\begin{equation} \label{eq:main_d_oper_from_d}
    D = \dot{X}_i \Bigg[  
    \sum_{k=0}^{\infty} \frac{(-1)^k}{(k+1)!} \mbf{Q}^k
    \Bigg]_{ij} E_j
    \equiv \dot{X}_i d_{ij} E_j.
\end{equation}
We collect the elements $d_{ij}$ in a matrix $\mbf{D}$.
This matrix encodes the structure of the Lie algebra (through the structure constants) and the information tied to 
the specific element $X$ and provides, in essence, a local basis for expressing
$D$ directly in terms of the generators.

If we choose $s = X_k$, then it must be the case that $\dot{X}_i = \delta_{ki}$ and the
equation above simplifies to
\begin{equation} \label{eq:main_ds_oper_from_d}
    D = d_{kj} E_j
\end{equation}
In this case, we see that the $\mbf{D}$ matrix simply contains the expansion coefficients for the operator $D$.
In order to compute $\mbf{D}$, we assume that $\mbf{Q}$ is diagonalizable as
$\mbf{Q} = \mbf{P} \bm{\Lambda} \mbf{P}^{-1}$ and write
\begin{align} \label{eq:main_d_using_diag}
    \mbf{D}
    &= \mbf{P} \varphi( \bm{\Lambda} ) \mbf{P}^{-1}
\end{align}
where the function
\begin{align} \label{eq:main_func_phi_def}
    \varphi(z) 
    &= \sum_{k=0}^{\infty} \frac{(-1)^k}{(k+1)!} z^k
    = 
    \begin{dcases}[c]
        \displaystyle\frac{1 - \mrm{exp}(-z)}{z} & \text{if } z \neq 0 \\
        1                                        & \text{if } z =    0
    \end{dcases}
\end{align}
is applied to the diagonal elements $\lambda_i$ of $\bm{\Lambda}$, i.e. to the eigenvalues of $\mbf{Q}$.
We have used the limit $\varphi(z) \rightarrow 1$ as $z \rightarrow 0$ to
define $\varphi(0)$. 
Looking at Eq.~\eqref{eq:main_d_using_diag} it is evident that $\mbf{D}$ has eigenvalues 
$\varphi( \lambda_i )$. This implies that $\mbf{D}$ is invertible
when $\varphi( \lambda_i ) \neq 0$, i.e. when $\lambda_i \neq 2 \pi i k$ where $k$ is a non-zero integer.

\subsection{Simplified equations of motion} \label{ssec:simplified_eoms}

Section~\ref{ssec:similarity_transformed_derivative} provides a feasible and very general
procedure for computing the $\mathbf{D}^{m}$ matrices. However, it is not at all obvious that this
procedure will lead to efficient working equations.
In Appendix~\ref{appendix:d_oper_special} we work out the details and show
that an attractive result can indeed be obtained provided the matrix $\mbf{K}^m$ 
(which holds the elements of the
vector $\bm{\kappa}^m$) is diagonalizable. 
Dropping the mode index for clarity, we
assume that
\begin{align} \label{eq:kappa_diag_main_text}
    \mbf{K} &= \mbf{R} \bm{\mu} \mbf{L}^{\dagger}, \quad 
    \mbf{L}^{\dagger} \mbf{R} = \mbf{R} \mbf{L}^{\dagger} = \mbf{1} 
\end{align}
where $\bm{\mu}$ is diagonal and construct an auxiliary matrix $\mbf{\Omega}$ with elements
\begin{align} \label{eq:omega_element}
    \omega_{pq} = 1 / \varphi(\mu_p - \mu_q).
\end{align}
The numbers $\varphi(\mu_p - \mu_q)$ are in fact the eigenvalues of
$\mbf{D}^{\trans}$ and
our assumption that $\mbf{D}^{\trans}$ is invertible is equivalent to
assuming
\begin{align} \label{eq:DT_condition}
    \varphi(\mu_p - \mu_q) \neq 0 
    \quad \Longleftrightarrow \quad
    \mu_p - \mu_q \neq 2\pi i k
\end{align}
where
$k$ is a non-zero integer.
With these prerequisites in place, the result is given by
\begin{align} \label{eq:kappa_transform}
    \dot{\mbf{K}}
    = -i \mbf{R} \big( \mbf{\Omega} \circ \big(\mbf{L}^{\dagger} \mbf{G} \mbf{R} \big) \big) \mbf{L}^{\dagger}
\end{align}
where $\dot{\mbf{K}}$ and $\mbf{G}$ are the matrices holding the elements of the vectors $\dot{\bm{\kappa}}$
and $\mbf{g}$, respectively, for each mode (see Appendix~\ref{appendix:d_oper_special} for details).
The symbol $\circ$ denotes the Hadamard (or element-wise) product.
Although the transformation in Eq.~\eqref{eq:kappa_transform} may look somewhat unusual, it is not expensive to perform
and can be easily implemented.

In summary, the steps necessary for the exponential parametrization are as follows:
(i) solve Eq.~\eqref{eq:eom_kappa_tilde_dot} to obtain $\mbf{G}$; 
(ii) diagonalize $\mbf{K}$ as in Eq.~\eqref{eq:kappa_diag_main_text};
(iii) compute $\mathbf{\Omega}$ using Eq.~\eqref{eq:omega_element};
and (iv) obtain $\dot{\mbf{K}}$ from Eq.~\eqref{eq:kappa_transform}.
Regarding the computational cost, we note that the diagonalization step scales
as $\mathcal{O}(N^3)$, where $N$ is the dimension of $\mathbf{K}$, i.e. the number of basis functions.
The computation of Eq.~\eqref{eq:kappa_transform} involves $4N^3 + N^2$ multiplications
if the matrix and Hadamard products are done sequentially. In total,
the computational cost arising from the exponential parametrization scales as $\mathcal{O}(N^3)$
(keeping only the leading term).

In the vibrational case, the diagonalization of $\mathbf{K}$ and computation of Eq.~\eqref{eq:kappa_transform} 
is performed separately for each mode so the cost
is $\mathcal{O}(MN^3)$, where $M$ is the number of modes and $N$ is the number of basis functions
per mode. It is important to note that $N$ does not scale with the size of the system, 
so the
additional steps involved in the exponential parametrization
have a cost that is simply linear in $M$.
This cost can safely be considered negligible.

For electrons, the number of basis functions does scale with the size of the system
so the cost is not negligible as such.
\revision[MGH]{}{In addition, $N$ may conceivably be very large (e.g. in
the case of grid-based methods) so that diagonalization of $\mathbf{K}$
is not realistic. 
In cases where the diagonalization is feasible, the 
$\mathcal{O}(N^3)$ scaling is often much lower
}%
than the scaling arising from the wave function expansion
and from integral transformations.
In \revision[MGH]{}{those cases}, the
operations related to the basis set parametrization are
not decisive for the cost of evaluating the full set of \acp{eom}.

Finally, it should be recalled that the most important computational
difference between various choices of basis set parametrization may
derive from the performance of the numerical integration, i.e. from the number of \ac{eom} evaluations.
For a given choice of integration algorithm, this number is determined by the average integrator step size,
which we will consider in Sec.~\ref{sec:results}. 
\subsection{Double exponential parametrization} \label{ssec:double_exp}
For each mode we may choose to parametrize the \revision[MGH]{}{basis functions} in terms of the unique polar decomposition\cite{hallLieGroupsLie2015},
\begin{align} \label{eq:polar_decomp}
    e^{\revision[MGH]{}{\hat{\kappa}}} = e^{\revision[MGH]{}{\hat{\kappa}}'} e^{\revision[MGH]{}{\hat{\kappa}}''}
\end{align}
where $\revision[MGH]{}{\hat{\kappa}}'$ is anti-Hermitian and $\revision[MGH]{}{\hat{\kappa}}''$ is Hermitian (we have dropped the mode index for clarity).
These properties imply that $e^{\revision[MGH]{}{\hat{\kappa}}'}$ is unitary while $e^{\revision[MGH]{}{\hat{\kappa}}''}$ is positive definite (i.e. Hermitian with
strictly positive eigenvalues). We note that double exponential orbital transformations have been considered
by Olsen\cite{olsenNovelMethodsConfiguration2015} in the context of ground state \ac{ci} calculations with non-orthogonal
orbitals.

In order to motivate such a parametrization we need to consider the situation
where the basis set is divided into active and secondary subsets, with the wave function
being expanded in the active basis alone. The exponential parametrization employed in the present
work ensures that the bra basis functions (given by $e^{-\revision[MGH]{}{\hat{\kappa}}}$) and the ket basis functions
(given by $e^{\revision[MGH]{}{\hat{\kappa}}}$) are biorthonormal by construction. Specifically, the
active bra and ket functions are biorthonormal. However, there is no guarantee
that they span the same space. Although this situation
is allowed by the formalism, we have found\cite{hojlundBivariationalTimedependentWave2022}
that the active bra and ket basis functions sometimes tend to drift very far apart,
even to an extent that eventually causes numerical breakdown at long integration times. In order to alleviate this
issue, we converted the linear basis set parametrization to a parametrization based on polar decomposition,
which allowed a (non-variational) restriction that was shown to enhance numerical stability by locking the bra and ket spaces together.
The linear and polar parametrizations of Ref.~\citenum{hojlundBivariationalTimedependentWave2022}
are exactly equivalent to the single exponential and double exponential parametrizations of
the present work. In particular, the double exponential parametrization allows a restriction
analogus to that of the restricted polar parametrization. We do not consider this restriction explicitly in
the present work since it is easily introduced by setting appropriate matrix elements to zero
(see Ref.~\citenum{hojlundBivariationalTimedependentWave2022} for details).

Returning to the derivations, we note that Eq.~\eqref{eq:polar_decomp} implies that
\begin{align} \label{eq:polar_decomp_time_deriv}
    e^{-\revision[MGH]{}{\hat{\kappa}}} \dv{e^{\revision[MGH]{}{\hat{\kappa}}}}{t}  = e^{-\revision[MGH]{}{\hat{\kappa}}''} e^{-\revision[MGH]{}{\hat{\kappa}}'} \dv{e^{\revision[MGH]{}{\hat{\kappa}}'} e^{\revision[MGH]{}{\hat{\kappa}}''}}{t},
\end{align}
which we will use as our starting point. Rather than rederiving everything from scratch, we will
seek to convert the \acp{eom} derived so far to the double exponential format. We start by noting that
the left-hand side of Eq.~\eqref{eq:polar_decomp_time_deriv} can be written as
\begin{align} \label{eq:polar_decomp_time_deriv_left}
    e^{-\revision[MGH]{}{\hat{\kappa}}} \dv{e^{\revision[MGH]{}{\hat{\kappa}}}}{t} 
    &= \sum_{pq} \dot{\kappa}_{pq} \, e^{-\revision[MGH]{}{\hat{\kappa}}} \pdv{e^{\revision[MGH]{}{\hat{\kappa}}}}{\kappa_{pq}} \nn
    &= \sum_{pq} \dot{\kappa}_{pq} D_{pq} \nn
    &= \sum_{pq} \sum_{rs} \dot{\kappa}_{pq} d_{(pq)(rs)} E_{rs} \nn
    &= -i\sum_{rs} g_{rs} E_{rs} \nn
    &= -i \revision[MGH]{}{\hat{g}}
\end{align}
where we have used Eqs.~\eqref{eq:d_oper_def} and \eqref{eq:kappa_tilde_dot_def}.
The right-hand side is equal to
\begin{align} \label{eq:polar_decomp_time_deriv_right}
    e^{-\revision[MGH]{}{\hat{\kappa}}''} e^{-\revision[MGH]{}{\hat{\kappa}}'} \dv{e^{\revision[MGH]{}{\hat{\kappa}}'} e^{\revision[MGH]{}{\hat{\kappa}}''}}{t}
    &=  e^{-\revision[MGH]{}{\hat{\kappa}}''} \Big( e^{-\revision[MGH]{}{\hat{\kappa}}'} \dv{e^{\revision[MGH]{}{\hat{\kappa}}'}}{t} \Big) e^{\revision[MGH]{}{\hat{\kappa}}''} 
    + \Big( e^{-\revision[MGH]{}{\hat{\kappa}}''} \dv{e^{\revision[MGH]{}{\hat{\kappa}}''}}{t} \Big) \nn
    &= e^{-\revision[MGH]{}{\hat{\kappa}}''} (-i \revision[MGH]{}{\hat{g}}') e^{\revision[MGH]{}{\hat{\kappa}}''} 
    + (- i \revision[MGH]{}{\hat{g}}'')
\end{align}
with the definitions
\begin{subequations} \label{eq:kappa_tilde_dot_prime_and_dprime_def}
\begin{align}
    \revision[MGH]{}{\hat{g}}'  &= i e^{-\revision[MGH]{}{\hat{\kappa}}'} \dv{e^{\revision[MGH]{}{\hat{\kappa}}'}}{t}, \\
    \revision[MGH]{}{\hat{g}}'' &= i e^{-\revision[MGH]{}{\hat{\kappa}}''} \dv{e^{\revision[MGH]{}{\hat{\kappa}}''}}{t}.
\end{align}
\end{subequations}
It follows from Eqs.~\eqref{eq:kappa_tilde_dot_prime_and_dprime_def} and the properties of $\revision[MGH]{}{\hat{\kappa}}'$ and $\revision[MGH]{}{\hat{\kappa}}''$ that
\begin{subequations} \label{eq:properties_of_kappa_tilde_dot_prime_and_dprime_def}
\begin{align}
    (\revision[MGH]{}{\hat{g}}')^{\dagger}  &= \revision[MGH]{}{\hat{g}}',  \\
    (\revision[MGH]{}{\hat{g}}'')^{\dagger} &= e^{\revision[MGH]{}{\hat{\kappa}}''}  (-\revision[MGH]{}{\hat{g}}'') e^{-\revision[MGH]{}{\hat{\kappa}}''} \label{eq:property_of_g_dprime}
\end{align}
\end{subequations}
which through Eqs.~\eqref{eq:polar_decomp}--\eqref{eq:polar_decomp_time_deriv_right}
implies
\begin{align}
    \revision[MGH]{}{\hat{g}} = e^{-\revision[MGH]{}{\hat{\kappa}}''} \big( \revision[MGH]{}{\hat{g}}' -  (\revision[MGH]{}{\hat{g}}'')^{\dagger}   \big) e^{\revision[MGH]{}{\hat{\kappa}}''}.
\end{align}
In matrix notation, this is
\begin{align} \label{eq:kappa_tilde_related_to_prime_and_dprime}
    \mbf{G}
    &= e^{-\mbf{K}''} \big( \mbf{G}' -  (\mbf{G}'')^{\dagger}   \big) e^{\mbf{K}''} \nn
    &= \mbf{P}^{-1} \big( \mbf{G}' -  (\mbf{G}'')^{\dagger}   \big) \mbf{P}
\end{align} 
where we have introduced $\mbf{P} = e^{\mbf{K}''}$
for notational convenience.
Since we have taken the primitive basis to be orthonormal (see Sec.~\ref{ssec:parametrization}), the properties
of the operators $\revision[MGH]{}{\hat{g}}'$ and $\revision[MGH]{}{\hat{g}}''$ carry over directly to their matrix representations (we may for example
conclude that the matrix $\mbf{G}'$ is Hermitian since the operator $\revision[MGH]{}{\hat{g}}'$ is Hermitian).
Equation~\eqref{eq:kappa_tilde_related_to_prime_and_dprime} has exactly the same kind of 
structure that was encountered in Ref. \citenum{hojlundBivariationalTimedependentWave2022}. Similar to that work, we define a similarity transformed version of
$\mbf{G}$, i.e.
\begin{align} \label{eq:G_similarity_transformed}
    \bar{\mbf{G}} = \mbf{P} \mbf{G} \mbf{P}^{-1} = \mbf{G}' -  (\mbf{G}'')^{\dagger},
\end{align} 
and use the Hermitianity of $\mbf{G}'$ to write
\begin{subequations} \label{eq:hermitian_and_antihermitian_parts_of_G_sim}
    \begin{align}
        \mathbb{H}(\bar{\mbf{G}}) &= \mbf{G}' - \mathbb{H}(\mbf{G}''), \label{eq:hermitian_and_antihermitian_parts_of_G_sim_GV} \\
        \mathbb{A}(\bar{\mbf{G}}) &= \mathbb{A}(\mbf{G}'')             \label{eq:hermitian_and_antihermitian_parts_of_G_sim_GP}
    \end{align}
\end{subequations}
where $\mathbb{H}$ and $\mathbb{A}$ denote the Hermitian and anti-Hermitian parts, respectively, of a square matrix.
In order to progress we combine $\mbf{G}'' = \mathbb{A}(\mbf{G}'') + \mathbb{H}(\mbf{G}'')$
with the property from Eq.~\eqref{eq:property_of_g_dprime} to get the following equation:
\begin{align} \label{eq:sylvester}
    \mathbb{A}(\mbf{G}'')   \mbf{P}  - \mbf{P} \, \mathbb{A}(\mbf{G}'')   &= \mathbb{H}(\mbf{G}'') \mbf{P} + \mbf{P} \, \mathbb{H}(\mbf{G}'').
\end{align} 
The matrix $\mbf{P} = e^{\mbf{K}''}$ is known from the outset while $\mathbb{A}(\mbf{G}'')$ is given by 
Eq.~\eqref{eq:hermitian_and_antihermitian_parts_of_G_sim_GP} and so the left-hand side of Eq.~\eqref{eq:sylvester} is
known. Equations such as Eq.~\eqref{eq:sylvester} are called Lyapunov or, more generally, Sylvester equations and are well-known in the 
mathematical literature.\cite{laubMatrixAnalysisScientists2005} We showed in the appendix of Ref. \citenum{hojlundBivariationalTimedependentWave2022} that
Eq.~\eqref{eq:sylvester} can be efficiently solved for $\mathbb{H}(\mbf{G}'')$ provided
that an eigenvalue decomposition of $\mbf{P}$ is available.
Before stating the result, we diagonalize $\mbf{K}'$ (anti-Hermitian) and $\mbf{K}''$ (Hermitian) as
\begin{subequations} \label{eq:Kp_Kpp_diag}
    \begin{gather}
        \mbf{K}'  = \mbf{S} \bm{\eta}  \mbf{S}^{\dagger}, \quad \mbf{S}^{\dagger} \mbf{S}  =  \mbf{S} \mbf{S}^{\dagger} = \mbf{1},  \label{eq:Kp_diag}\\
        \mbf{K}'' = \mbf{T} \bm{\zeta} \mbf{T}^{\dagger}, \quad \mbf{T}^{\dagger} \mbf{T}  =  \mbf{T} \mbf{T}^{\dagger} = \mbf{1}.  \label{eq:Kpp_diag}
    \end{gather}
\end{subequations}
The eigenvalues $\eta_p$ are purely imaginary while the $\zeta_p$ are purely real.
Equation~\eqref{eq:Kpp_diag} allows us to write
\begin{align}
    \mbf{P} &= \exp(\mbf{K}'')= \mbf{T} \exp(\bm{\zeta}) \mbf{T}^{\dagger} 
    = \mbf{T} \bm{\epsilon} \mbf{T}^{\dagger}
\end{align}
where the eigenvalues $\epsilon_p = \exp(\zeta_p)$, are purely real and strictly positive.
The expression for $\mathbb{H}(\mbf{G}'')$ now reads
\begin{align} \label{eq:P_generator_hermitian_part}
    \mathbb{H}(\mbf{G}'') = \mbf{T} \big( \mbf{\Gamma} \circ \big(\mbf{T}^{\dagger} \mathbb{A}(\bar{\mbf{G}}) \mbf{T} \big) \big) \mbf{T}^{\dagger}
\end{align}
where the matrix $\mbf{\Gamma}$ has elements
\begin{align} \label{eq:gamma_element}
    \gamma_{pq} = \frac{-\epsilon_p + \epsilon_q}{\epsilon_p + \epsilon_q}.
\end{align}
Note that $\mbf{\Gamma}$ is real anti-symmetric and thus anti-Hermitian by construction. The denominator
in Eq.~\eqref{eq:gamma_element} is always greater than zero since $\epsilon_p > 0$ as already mentioned.
Combining Eqs.~\eqref{eq:hermitian_and_antihermitian_parts_of_G_sim} and \eqref{eq:P_generator_hermitian_part}
now yields
\begin{subequations} \label{eq:VP_generators_from_terms}
    \begin{alignat}{2}
        \mbf{G}'  &= \mathbb{H}(\bar{\mbf{G}}) &&+ \mbf{T} \big( \mbf{\Gamma} \circ \big(\mbf{T}^{\dagger} \mathbb{A}(\bar{\mbf{G}}) \mbf{T} \big) \big) \mbf{T}^{\dagger},  \label{eq:V_generator_from_terms} \\
        \mbf{G}'' &= \mathbb{A}(\bar{\mbf{G}}) &&+ \mbf{T} \big( \mbf{\Gamma} \circ \big(\mbf{T}^{\dagger} \mathbb{A}(\bar{\mbf{G}}) \mbf{T} \big) \big) \mbf{T}^{\dagger}.  \label{eq:P_generator_from_terms}
    \end{alignat}
\end{subequations}
Having determined $\mbf{G}'$ and $\mbf{G}''$ we recover $\dot{\mbf{K}}'$ and $\dot{\mbf{K}}''$ through transformations analogous to
Eq.~\eqref{eq:kappa_transform}. 
For that purpose we construct
the auxiliary matrices $\mbf{\Omega}'$ and $\mbf{\Omega}''$ with
elements
\begin{subequations}
    \begin{align}
        \omega_{pq}'  &= 1/\varphi(\eta_p  - \eta_q ), \\
        \omega_{pq}'' &= 1/\varphi(\zeta_p - \zeta_q).
    \end{align}
\end{subequations}
The eigenvalues $\eta_p$ and $\zeta_p$ are taken from Eqs.~\eqref{eq:Kp_Kpp_diag} and the function $\varphi$
is given by Eq.~\eqref{eq:main_func_phi_def}.
The resulting time derivatives then follow as
\begin{subequations} \label{eq:Kp_Kpp_dot}
    \begin{align}
        \dot{\mbf{K}}'  &= -i \mbf{S} \big( \mbf{\Omega}'  \circ \big(\mbf{S}^{\dagger} \mbf{G}'  \mbf{S} \big) \big) \mbf{S}^{\dagger}, \label{eq:kappa_prime_dot} \\
        \dot{\mbf{K}}'' &= -i \mbf{T} \big( \mbf{\Omega}'' \circ \big(\mbf{T}^{\dagger} \mbf{G}'' \mbf{T} \big) \big) \mbf{T}^{\dagger}. \label{eq:kappa_double_prime_dot}
    \end{align}
\end{subequations}
The matrix $\mbf{K}''$ has real eigenvalues and so the condition in Eq.~\eqref{eq:DT_condition} always holds. Thus Eq.~\eqref{eq:kappa_double_prime_dot}
is never singular. Conversely, the matrix $\mbf{K}'$ has purely imaginary eigenvalues so Eq.~\eqref{eq:kappa_prime_dot} can become singular.

The overall procedure for the double exponential parametrization is as follows: (i) solve
Eq.~\eqref{eq:eom_kappa_tilde_dot} to obtain $\mbf{G}$;
(ii) diagonalize $\mbf{K}'$ and $\mbf{K}''$ as in Eqs.~\eqref{eq:Kp_Kpp_diag}; 
(iii) compute $\bar{\mbf{G}}$ in Eq.~\eqref{eq:G_similarity_transformed}; 
(iv) compute $\mbf{G}'$ and $\mbf{G}''$ in Eqs.~\eqref{eq:VP_generators_from_terms}; 
and (v) recover $\dot{\mbf{K}}'$ and $\dot{\mbf{K}}''$
from Eqs.~\eqref{eq:Kp_Kpp_dot}.
The discussion of computational cost is completely 
analogous to that at the end of Sec.~\ref{ssec:simplified_eoms}.
 
\subsection{Removing singularities} \label{ssec:singularities}
Before implementing the single and double exponential \revision[MGH]{}{basis set} \acp{eom}, we need to
adress the possible singularities. We first note that one is not likely to encounter
exact singularities in a numerical settings but rather near-singularities. Such near-singularities
will not necessarily cause numerical breakdown but they will result in very large $\dot{\mbf{K}}^m$
(single exponential case)
or $\dot{\mbf{K}}^{\prime m}$ (double exponential case), thus making the \acp{eom} hard to propagate.
We expect that a good, adaptive integrator will be able to manage such difficulties at the price
of temporarily reducing the step size. Although such a situation is not disastrous it should
be avoided since the cost of propagating the wave function is inversely proportional to the
average step size.

For the single exponential case, we remove a (near-)singularity in a given mode $m$ by absorbing the non-zero $\revision[MGH]{}{\hat{\kappa}}^m$
into the Hamiltonian:
\begin{align}
    H \leftarrow \exp(-\revision[MGH]{}{\hat{\kappa}}^m) H \exp(\revision[MGH]{}{\hat{\kappa}}^m)
\end{align}
In practice this simply amounts to transforming the Hamiltonian integrals and resetting $\revision[MGH]{}{\hat{\kappa}}^m$ to zero
(equivalently, this process can be phrased as a transformation of the primitive basis).
The same procedure applies to $\revision[MGH]{}{\hat{\kappa}}^{\prime m}$ in the double exponential case and we note
that a transformation by the unitary operator $\exp(\revision[MGH]{}{\hat{\kappa}}^{\prime m})$ keeps the primitive
basis orthonormal (which was needed for the derivations in Sec.~\ref{ssec:double_exp}).

As a criterion for resetting mode $m$,
we take the simple approach of monitoring the numbers
$ \varphi_{p^m q^m}^m = \varphi(\mu_{p^m}^m - \mu_{q^m}^m)$,
which in the single exponential case are simply the eigenvalues of $(\mbf{D}^{m})^{\trans}$; see Eqs.~\eqref{eq:main_func_phi_def}
and \eqref{eq:DT_condition}.
We currently perform the reset of mode $m$ if
\begin{align} \label{eq:reset_criterion}
    | \varphi^m_\mrm{min} | = \min_{p^m q^m}  \big|  \varphi_{p^m q^m}^m \big| < \tau
\end{align}
where $\tau$ is a user-defined threshold. The minimum eigenvalue has the benefit of not scaling significantly
with the size of the primitive basis in contrast to e.g. the determinant of $(\mbf{D}^{m})^{\trans}$ (which
is the product of the eigenvalues).

It should be noted that algorithms for integrating \acp{ode} typically
evaluate the \acp{eom} several times in order to determine an appropriate step. We perform the
check in Eq.~\eqref{eq:reset_criterion} after each such evaluation but the reset is only
done after the step is completed and if the criterion was fulfilled at least once.
Choosing the threshold $\tau$ sufficiently large results in the reset of every mode after
every integrator step.  
\subsection{Comparison to local derivatives} \label{ssec:local_derivative}
Section~\ref{ssec:singularities} describes how one can reset non-zero $\revision[MGH]{}{\hat{\kappa}}^m$
in order to avoid (near-)singularities in 
the \acp{eom}. Having $\revision[MGH]{}{\hat{\kappa}}^m = 0$ also results in simpler working equations, e.g.
\begin{align}
    \dot{\mbf{K}}^m = -i \mbf{G}^m ,
\end{align}
and one could thus consider an approach where the resets are performed before every
evaluation of the \acp{eom}, thus basing the propagation on temporally local
derivatives of the Lagrangian.
This is indeed the approach taken by
Kristiansen et al.\cite{kristiansenLinearNonlinearOptical2022} in their derivation
of (electronic) time-dependent orbital-optimized coupled cluster with double excitations
(TDOCCD), non-orthogonal TDOCCD (TDNOCCD) and the corresponding second-order approximations
TDOMP2 and TDNOMP2.

Although formally equivalent, it is not immediately clear to us that
the two approaches are fully equivalent in a numerical setting where time is discretized. 
To illustrate the point, consider the situation where initially $\revision[MGH]{}{\hat{\kappa}}^m = 0$ for each mode.
For the two sets of \acp{eom} to be completely equivalent, we require that they
result in the same integrator step, i.e. they should describe the same physical
evolution of the system. For definiteness, we consider a general (explicit or implicit)
Runge-Kutta method\cite{hairerSolvingOrdinaryDifferential2009b,hairerGeometricNumericalIntegration2006}:
\begin{subequations} \label{eq:RK_def}
    \begin{align}
        \mbf{y}_{n+1} &= \mbf{y}_{n} + h \sum_{i=1}^{s} b_i \mbf{k}_i \label{eq:RK_step} \\
        \mbf{k}_i &= f(t_n + c_i h, \mbf{y}_n + h \sum_{j=1}^{s} a_{ij} \mbf{k}_j), \quad i = 1, \ldots, s \label{eq:RK_eqs}
    \end{align}
\end{subequations}
Here, $\mbf{y}_n$ and $\mbf{y}_{n+1}$ are the vectors containing current and updated parameters, $h$ is the step size,
the scalars $c_i$, $b_i$ and $a_{ij}$ are the parameters defining the method, $s$ is the number of stages
and $f(t, \mbf{y}) = \dot{\mbf{y}}(t, \mbf{y})$.
The auxiliary vectors $\mbf{k}_i$ should satisfy the (generally non-linear) equations in Eq.~\eqref{eq:RK_eqs}.
Since the two approaches result in numerically different values of $f(t, \mbf{y})$, it seems to us
that the solution to Eq.~\eqref{eq:RK_eqs} and thus the integrator step predicted by Eq.~\eqref{eq:RK_step}
will differ (also when the $\revision[MGH]{}{\hat{\kappa}}^m$ are initially equal to zero).
We hypothesize that this difference is small when $h$ is small. 
Thus, this work does not question the usefullness of the local derivative approach. 
Rather, the point is that while the general formulation for non-zero $\revision[MGH]{}{\hat{\kappa}}^m$
might seem very complicated at the outset, it turns out that it is in fact
simple to implement 
with limited computational cost as discussed in Sec.~\ref{ssec:simplified_eoms}. 

\section{Implementation} \label{sec:implementation}
The single and double exponential modal \acp{eom} have been implemented
in the \ac{midas}\cite{artiukhinMidasCpp2022} in conjunction with the \ac{tdmvcc}\cite{madsenTimedependentVibrationalCoupled2020}
method. The existing \ac{tdmvcc} code has been refactored to make it largely agnostic to the
parametrization of the modals. Modal parameters are now stored in an abstract class that
is easily specialized using C++ class inheritance.
We note that the computationally intensive parts of the code are independent of the choice
of modal parametrization.

This modular design is possible since the time evolution of the modals
is always governed by the same set of eqations, Eq.~\eqref{eq:eom_kappa_tilde_dot}.
Having solved these equations (which is a mayor computational task),
the additional steps leading to linear\cite{madsenTimedependentVibrationalCoupled2020},
polar\cite{hojlundBivariationalTimedependentWave2022} or exponential parametrization (this work)
are all simple and cheap to perform. The computational overhead resulting from the
exponential parametrization (which requires the diagonalization of one-mode quantities
and a number of one-mode transformations) scales linearly with respect to the number of
modes and is negligible for all but the smallest systems.

We note that the \ac{tdmvcc} method is presently implemented using the general but inefficient \ac{fsmr}
framework that was introduced in Ref.~\citenum{hansenExtendedVibrationalCoupled2020}. This limits
our calculations to small systems (approximately six modes and rather small active basis sets).
Efficient, polynomial-scaling implementations of TDMVCC are the subject of current research in our group. 

\section{Numerical examples} \label{sec:results}
\subsection{Computational details}

We consider two numerical examples from Ref. \citenum{madsenTimedependentVibrationalCoupled2020}
in order to study the performance and
stability of the single and double exponential modal parametrizations.
The first example is the \ac{ivr} of water after the excitation
of the symmetric stretch to $n = 2$. The initial state is obtained
as the $[0, 2, 0]$ state on the harmonic part of the \ac{pes}, i.e.
the initial state is a simple harmonic oscillator state. The wave
packet is then propagated at the TDMVCC$[2]$ level on the full (anharmonic and coupled) \ac{pes}
using the \ac{dop853} integrator\cite{hairerSolvingOrdinaryDifferential2009b}
with integrator tolerances $\tau_\mrm{abs} = 10^{-10}$ and $\tau_\mrm{rel} = 10^{-10}$.
The calculation uses 30 primitive modals and 4 active modals for each mode.

The second example is the Franck--Condon emission ($S_1 \rightarrow S_0$) of the 5D \textit{trans}-bithiophene
model of Ref.~\citenum{madsenVibrationallyResolvedEmission2019}. The initial state is taken to be the \ac{vscf}
ground state of the $S_1$ electronic surface. The wave packet is then placed
on the $S_0$ surface and propagated at the TDMVCC$[2]$ level using the same integrator settings as above. Once again,
we use 30 primitive and 4 active modals for each mode.

For both examples we repeat the calculations using linearly parametrized modals
(see Refs.~\citenum{madsenTimedependentVibrationalCoupled2020} and \citenum{hojlundBivariationalTimedependentWave2022})
and with single and double exponentially parametrized modals. Reset thresholds $\tau = 0, 0.1, 0.3, 0.5, 0.7, \infty$ are
considered. The threshold $\tau = 0$ corresponds to never resetting the
\revision[MGH]{}{basis set parameters,}
while the
threshold $\tau = \infty$ corresponds to resetting after every step.

In both cases we show \acp{acf} to illustrate the equivalence of the various modal
parametrizations. The \ac{acf} is not an observable as such
but is conveys important information about the dynamics of the system while being sensitive
to errors in the wave function parameters.

\subsection{\Aclp{acf}}
\begin{figure}[H]
    \centering
    \includegraphics[width=0.74\columnwidth]{{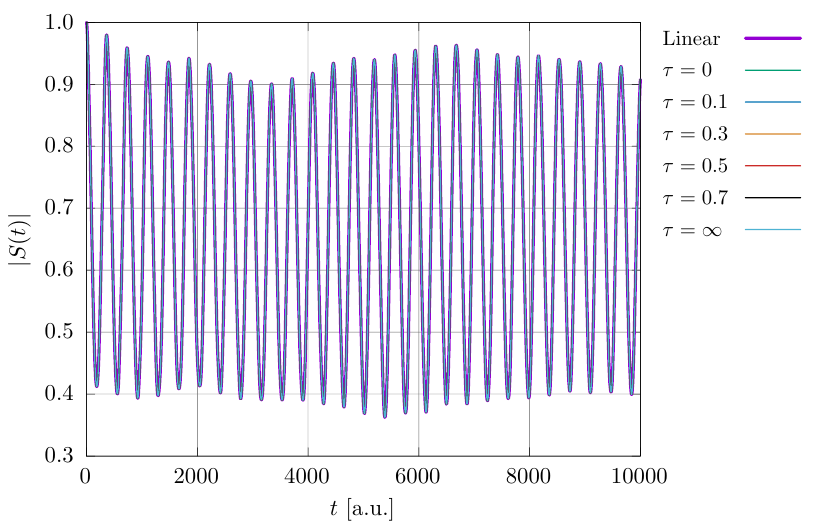}}
    \caption{\Acp{acf} for the \ac{ivr} of water described
    at the TDMVCC$[2]$ level. The reference calculation (purple) uses linearly parametrized modals,
    while the remaining calculations use single exponentially parametrized modals with a
    number of reset thresholds $\tau$ (see text).
    Note that the lines coincide so some are not visible.}
    \label{fig:water_tdmvcc2_autocorr}
\end{figure}

Figure~\ref{fig:water_tdmvcc2_autocorr} shows \acp{acf} for the linear and single exponential
water \ac{ivr} calculations. The \acp{acf} all coincide perfectly, demonstrating the
numerical equivalence of the various modal parametrizations. The dynamics is strongly oscillatory
with a dominating period of roughly \SI{370}{{a.u.}} (\SI{9}{fs}).

\begin{figure}[H]
    \centering
    \includegraphics[width=0.74\columnwidth]{{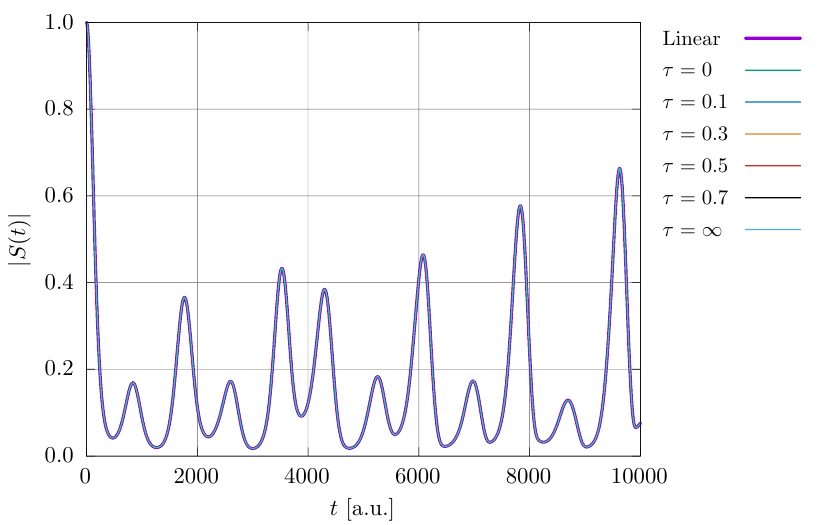}}
    \caption{\Acp{acf} for the 5D \textit{trans}-bithiophene model described
    at the TDMVCC$[2]$. The reference calculation (purple) uses linearly parametrized modals,
    while the remaining calculations use double exponentially parametrized modals with a
    number of reset thresholds $\tau$ (see text).
    Note that the lines coincide so some are not visible.}
    \label{fig:tbithio_tdmvcc2_autocorr}
\end{figure}

Figure~\ref{fig:tbithio_tdmvcc2_autocorr} display \acp{acf} for the 5D \textit{trans}-bithiophene model
using linearly and double exponentially parametrized modals. We once again observe perfect agreement
between the different parametrizations. This time the dynamics is significantly slower, which
is also reflected in the integrator step size (see later).

\subsection{Integrator step size}

\begin{figure}[H]
    \centering
    \includegraphics[width=0.8\columnwidth]{{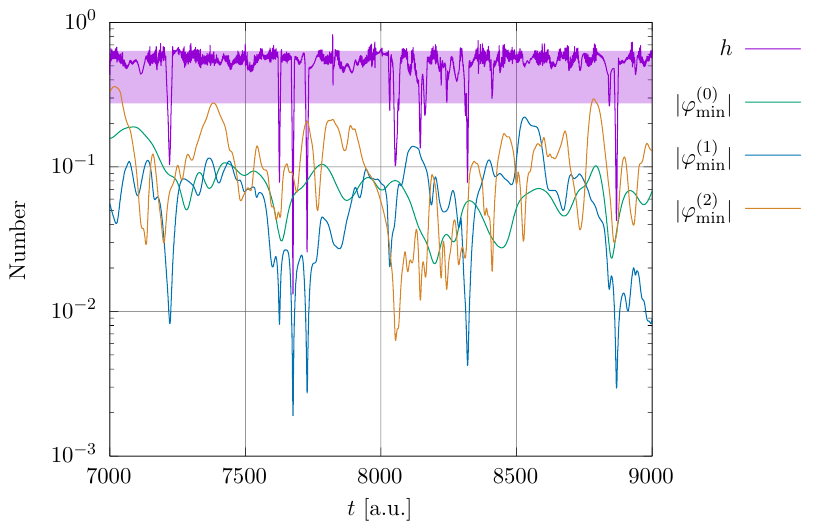}}
    \caption{Integrator step size ($h$) and $|\varphi^m_\mrm{min}|$ for the \ac{ivr} of water described
    at the TDMVCC$[2]$ level using single exponentially parametrized modals ($\tau = 0$, i.e. no resets).
    The shaded area covers the mean step size plus/minus its standard deviation (computed within the shown time interval).
    Only part of the full time interval is shown for greater visibility.}
    \label{fig:water_tdmvcc2_exp_stepsize}
\end{figure}

\begin{figure}[H]
    \centering
    \includegraphics[width=0.8\columnwidth]{{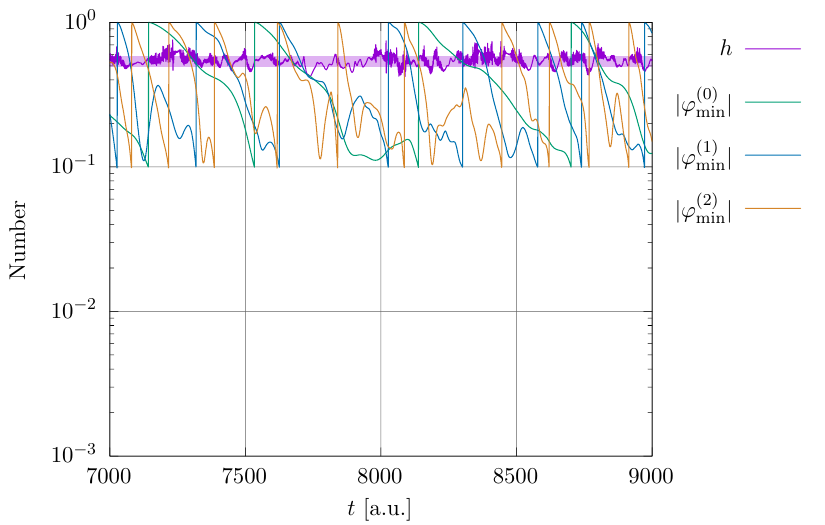}}
    \caption{Integrator step size ($h$) and $|\varphi^m_\mrm{min}|$ for the \ac{ivr} of water described
    at the TDMVCC$[2]$ level using single exponentially parametrized modals ($\tau = 0.1$).
    The shaded area covers the mean step size plus/minus its standard deviation (computed within the shown time interval).
    Only part of the full time interval is shown for greater visibility.}
    \label{fig:water_tdmvcc2_exp_trans1.0e-1_stepsize}
\end{figure}

Having demonstrated the equivalence of the linear and exponential formalisms
(and the correctness of the implementation) we turn to examine their performance
and behaviour with respect to the integrator step size. 
Figure~\ref{fig:water_tdmvcc2_exp_stepsize} shows the step size and
the minimum eigenvalues $|\varphi^m_\mrm{min}|$ of Eq.~\eqref{eq:reset_criterion}
for the water \ac{ivr} calculation using single exponentially parametrized modals
and $\tau = 0$ (i.e. no resets). The step size exhibits a series of very sharp dips that
coincide with small values of $|\varphi^m_\mrm{min}|$. This is exactly the kind of near-singularity
behaviour that was described in Sec.~\ref{ssec:singularities}. In this particular case, it seems
that mode $1$ (i.e. the symmetric stretch, which was initially excited) is responsible for
most near-singularities in the time interval shown. We note that the near-singularities
are very short-lived and that the integrator is perfectly capable of stepping through them.
This suggests that the formal existence of singularities in the modal \acp{eom} is not catastrophic
in a numerical setting. In Fig.~\ref{fig:water_tdmvcc2_exp_trans1.0e-1_stepsize}, the water \ac{ivr} calculation is repeated with $\tau = 0.1$.
The resets are clearly visible in the figure and have the effect of removing any (near-)singularities.
As a consequence, the step size is stabilized close to its mean value.

\begin{figure}[H]
    \centering
    \includegraphics[width=0.8\columnwidth]{{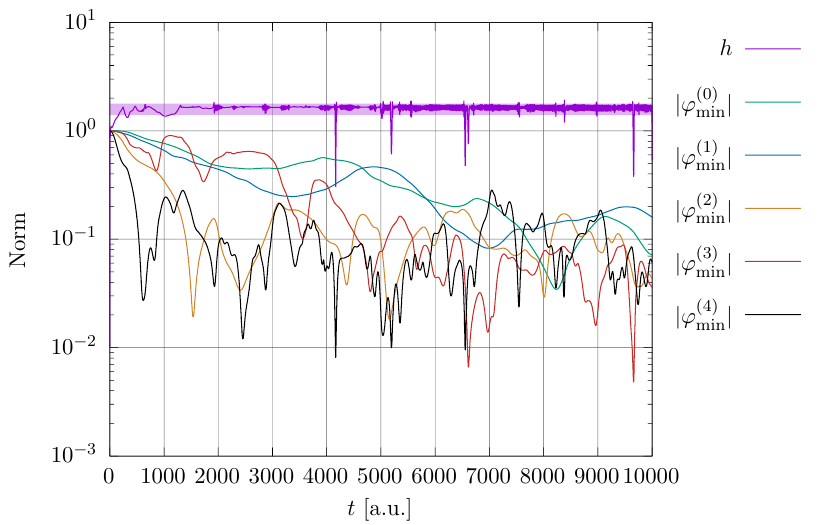}}
    \caption{Integrator step size ($h$) and $|\varphi^m_\mrm{min}|$ for the 5D \textit{trans}-bithiophene model described
    at the TDMVCC$[2]$ level using double exponentially parametrized modals ($\tau = 0$, i.e. no resets).
    The shaded area covers the mean step size plus/minus its standard deviation.}
    \label{fig:tbithio_tdmvcc2_double_exp_stepsize}
\end{figure}

\begin{figure}[H]
    \centering
    \includegraphics[width=0.8\columnwidth]{{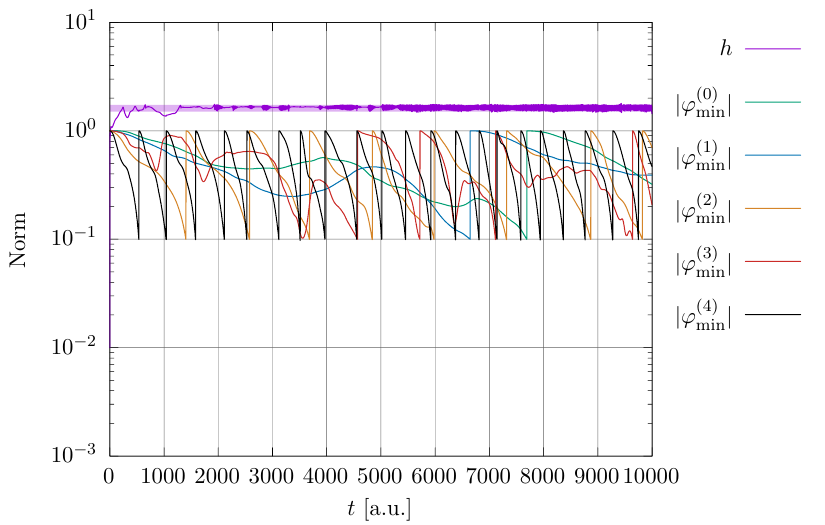}}
    \caption{Integrator step size ($h$) and $|\varphi^m_\mrm{min}|$ for the 5D \textit{trans}-bithiophene model described
    at the TDMVCC$[2]$ level using double exponentially parametrized modals ($\tau = 0.1$).
    The shaded area covers the mean step size plus/minus its standard deviation.}
    \label{fig:tbithio_tdmvcc2_double_exp_trans1.0e-1_stepsize}
\end{figure}

Analogous results are shown for the 5D \textit{trans}-bithiophene model 
in Figs.~\ref{fig:tbithio_tdmvcc2_double_exp_stepsize} and
\ref{fig:tbithio_tdmvcc2_double_exp_trans1.0e-1_stepsize} (using double exponentially
parametrized modals). A few near-singularities
are also observed for this system but they are much less frequent compared to the 
water \ac{ivr} case (this difference can be explained partially by the fact that
the \textit{trans}-bithiophene dynamics is simply slower).
The near-singularities are again removed efficiently by the threshold-based resets.

Figures S1--S4 in the supplementary material show very similar behaviour for the \ac{ivr} of water
with double exponential modals and the 5D \textit{trans}-bithiophene model with single exponential modals.

In order to assess performance in a more quantitative fashion we present average step sizes ($h_\mrm{mean}$)
for water and \textit{trans}-bithiophene in Tables~\ref{tab:water_tdmvcc2} and \ref{tab:tbithio_tdmvcc2}.
The tables also include the mean stepsize relative to a reference calculation with linearly parametrized modals, i.e.
\begin{align}
    h_\mrm{rel} = \frac{h_\mrm{mean}}{h_\mrm{mean}^\mrm{linear}}.
\end{align}
For water (Table~\ref{tab:water_tdmvcc2}), the thresholds $\tau = 0$ (no resets)
and $\tau = \infty$ (reset after every step) perform slightly worse than the linear reference
for single as well as double exponential calculations.
Considering Fig.~\ref{fig:water_tdmvcc2_exp_stepsize}, it is perhaps not surprising that $\tau = 0$
leads to smaller average steps due to the near-singularities.
The remaining thresholds ($\tau = 0.1, 0.3, 0.5, 0.7$) all perform slightly better than
the linear reference, although the difference is small (on the order of a few percent).
The single and double exponential calculations perform essentially the same.

For \textit{trans}-bithiophene (Table~\ref{tab:tbithio_tdmvcc2}), all calculations
with exponentially parametrized modals perform slightly better than the reference except
for the single exponential calculation with $\tau = 0$. The double exponentially
parametrized modals result in slightly larger average steps compared to the
single exponential case, but the difference is again small.

\setlength{\tabcolsep}{12pt}
\begin{table}[H]
    \centering
    \caption{Mean and relative step sizes for the \ac{ivr} of water described
    at the TDMVCC$[2]$ level. $h_\mrm{mean}$ is given in atomic units and
    $h_\mrm{rel}$ is given relative to the linear case.}
    \bigskip
    \begin{tabular}{
    l
    S[table-format=1.1] 
    S[table-format=1.4] 
    S[table-format=1.4] 
    }
\toprule
    Type &     {$\tau$} &  {$h_\mathrm{mean}$} &   {$h_\mathrm{rel}$} \\ 
\midrule
  Linear &        {---} &               0.4088 &               1.0000 \\ 
\midrule
  Single &            0 &               0.3880 &               0.9491 \\ 
         &          0.1 &               0.4182 &               1.0229 \\ 
         &          0.3 &               0.4175 &               1.0211 \\ 
         &          0.5 &               0.4151 &               1.0154 \\ 
         &          0.7 &               0.4119 &               1.0075 \\ 
         &   {$\infty$} &               0.4080 &               0.9979 \\ 
\midrule
  Double &            0 &               0.3879 &               0.9488 \\ 
         &          0.1 &               0.4178 &               1.0220 \\ 
         &          0.3 &               0.4174 &               1.0210 \\ 
         &          0.5 &               0.4151 &               1.0154 \\ 
         &          0.7 &               0.4129 &               1.0099 \\ 
         &   {$\infty$} &               0.4079 &               0.9977 \\ 
 \bottomrule
\end{tabular}
\label{tab:water_tdmvcc2}%
\end{table}%

\setlength{\tabcolsep}{12pt}
\begin{table}[H]
    \centering
    \caption{Mean and relative step sizes for the 5D \textit{trans}-bithiophene model described
    at the TDMVCC$[2]$ level. $h_\mrm{mean}$ is given in atomic units and
    $h_\mrm{rel}$ is given relative to the linear case.}
    \bigskip
    \begin{tabular}{
    l
    S[table-format=1.1] 
    S[table-format=1.4] 
    S[table-format=1.4] 
    }
\toprule
    Type &     {$\tau$} &  {$h_\mathrm{mean}$} &   {$h_\mathrm{rel}$} \\ 
\midrule
  Linear &        {---} &               1.5704 &               1.0000 \\ 
\midrule
  Single &            0 &               1.5574 &               0.9917 \\ 
         &          0.1 &               1.6075 &               1.0236 \\ 
         &          0.3 &               1.6080 &               1.0240 \\ 
         &          0.5 &               1.6088 &               1.0245 \\ 
         &          0.7 &               1.6090 &               1.0246 \\ 
         &   {$\infty$} &               1.6100 &               1.0253 \\ 
\midrule
  Double &            0 &               1.5901 &               1.0126 \\ 
         &          0.1 &               1.6184 &               1.0306 \\ 
         &          0.3 &               1.6186 &               1.0308 \\ 
         &          0.5 &               1.6194 &               1.0313 \\ 
         &          0.7 &               1.6184 &               1.0306 \\ 
         &   {$\infty$} &               1.6202 &               1.0318 \\ 
 \bottomrule
\end{tabular}
\label{tab:tbithio_tdmvcc2}%
\end{table}%

\section{Summary and outlook} \label{sec:summary}
A general set of \acp{eom} for time-dependent wave functions with
exponentially parametrized biorthogonal basis sets has been derived
in a fully bivariational framework. The non-trivial connection
to the \acp{eom} for linearly parametrized basis sets was
elucidated, thus offering a unified perspective on the two approaches.
In particular, it was shown that
the computationally intensive parts are in fact
identical for the two kinds of parametrizations.
The exponential parametrization can thus be implemented
on top of existing code with \revision[MGH]{}{limited programming effort.}

Careful analysis of the equations showed the existence of a
well-defined set of singularities related to the eigenvalues
of the matrices containing the basis set parameters. It was
demonstrated how these singularities can be removed
in a controlled manner 
by a simple update of the Hamiltonian integrals.
The monitoring of singularities requires only quantities
that are computed in any case.

The exponential \acp{eom} were subsequently used as a starting point
for deriving \acp{eom} for a double exponential parametrization, 
thus providing
a separation of the basis set time evolution into a unitary part
and a part describing the deviation from unitarity. The transformations 
necessary to obtain the double exponential formulation again carry
negligible additional cost.

Finally, we presented numerical results for calculations on water
and a 5D \textit{trans}-bithiophene model. The calculations showed
that the single and double exponential parametrizations result in
slightly larger step sizes compared to linearly parametrized reference
calculations.
Although the effect on step size is minor,
our findings underline the fact that one should not necessarily consider
the exponential basis set parametrization as more costly compared to
the linear parametrization. We have argued that the additional
operations needed for the exponential parametrization are computationally
cheap, while an increase in the average itegrator step size leads directly
to fewer costly evaluations of the \acp{eom}. 

Although the present work focuses on bivariational wave functions, we note that
the mathematical machinery for converting between linearly and exponentially
parametrized basis sets is also applicable to other types of wave functions.
We thus anticipate that these results will be usefull in future treatments of 
both nuclear quantum dynamics and time-dependent electronic structure.

\section*{Supplementary material} \label{sec:supplementary_material}
The supplementary material contains additional figures related to the
water and 5D \textit{trans}-bithiophene calculations presented within the article.

\section*{Acknowledgements}
O.C. acknowledges support from the Independent Research Fund Denmark through grant number 1026-00122B.
Computations were performed at the Centre for Scientific Computing Aarhus (CSCAA).

\section*{Author declarations}
\subsection*{Conflict of Interest}
The authors have no conflicts to disclose.

\subsection*{Author Contributions}
\textbf{Mads Greisen Højlund}: 
Conceptualization (equal);
Data curation (lead);
Formal analysis (equal);
Investigation (lead);
Software (lead);
Visualization (lead);
Writing -- original draft (lead);
Writing -- review \& editing (equal).
\textbf{Alberto Zoccante}:
Conceptualization (equal);
Formal analysis (equal);
Writing -- review \& editing (equal).
\textbf{Ove Christiansen}:
Conceptualization (equal);
Formal analysis (equal);
Funding acquisition (lead);
Project administration (lead);
Supervision (lead);
Writing -- review \& editing (equal).

\section*{Data availability}
The data that supports the findings of this study are available within the article and its supplementary material.

\appendix

\section{Similarity transformed derivative: Specialization} \label{appendix:d_oper_special}
\setcounter{equation}{0}
\renewcommand{\theequation}{\ref{appendix:d_oper_special}\arabic{equation}}
Direct application of Eq.~\eqref{eq:main_ds_oper_from_d} shows that 
\begin{equation}
    D_{pq} = e^{-\revision[MGH]{}{\hat{\kappa}}} \pdv{e^{\revision[MGH]{}{\hat{\kappa}}}}{\kappa_{pq}} = 
    \sum_{r s} d_{(pq)(rs)} E_{rs}
\end{equation}
where we have reintroduced explicit summation and left out the mode 
index $m$ to avoid notational clutter. This is identical to Eq.~\eqref{eq:d_oper_def}
but now includes a concrete recipe for computing the $\mbf{D}$ matrix. 
In order to actually carry out the computation we need to consider the structure constants
describing the generators $E_{pq}$. By writing
\begin{align}
    [E_{pq}, E_{rs}] 
    &= \delta_{qr} E_{ps} - \delta_{ps} E_{rq} \\
    &= \sum_{\bar{r} \bar{s}} (
    \delta_{qr} \delta_{\bar{r}p} \delta_{\bar{s}s} -
    \delta_{ps} \delta_{\bar{r}r} \delta_{\bar{s}q}
    ) E_{\bar{r}\bar{s}} 
\end{align}
we see that
\begin{equation}
    f^{\bar{r}\bar{s}}_{(pq)(rs)} = (
        \delta_{qr} \delta_{\bar{r}p} \delta_{\bar{s}s} -
        \delta_{ps} \delta_{\bar{r}r} \delta_{\bar{s}q}
    ).
\end{equation}
The $\mbf{Q}$ matrix is now given by 
\begin{align}
    Q_{(rs)(\bar{r}\bar{s})} 
    &= \sum_{pq} \kappa_{pq} f^{\bar{r}\bar{s}}_{(pq)(rs)} \nn
    &= \sum_{pq} \kappa_{pq} \delta_{qr} \delta_{\bar{r}p} \delta_{\bar{s}s} 
    -  \sum_{pq} \kappa_{pq} \delta_{ps} \delta_{\bar{r}r} \delta_{\bar{s}q} \nn
    &= \kappa_{\bar{r}r} \delta_{\bar{s}s} 
    -  \kappa_{s\bar{s}} \delta_{\bar{r}r} \label{eq:M_matrix_element}
\end{align}
or, equivalently,
\begin{align} \label{eq:M_matrix_kron}
    \mbf{Q} 
    &= \mbf{K}^{\trans} \otimes \mbf{1} - \mbf{1} \otimes \mbf{K} \nn
    &= \mbf{K}^{\trans} \otimes \mbf{1} + \mbf{1} \otimes (-\mbf{K}) \nn
    &= \mbf{K}^{\trans} \oplus (-\mbf{K}).
\end{align}
Here, $\otimes$ denotes the Kronecker product while $\oplus$ denotes the so-called 
Kronecker sum. 
The simple structure in Eq.~\eqref{eq:M_matrix_kron}
allows us to analyse and manipulate the $\mbf{Q}$ matrix
in a convenient way. 
Let $\mbf{K}$ have eigenvalues $\mu_1, \ldots, \mu_N$
where $N$ is the order of the matrix. Then $-\mbf{K}$ has eigenvalues 
$-\mu_1, \ldots, -\mu_N$ while $\mbf{Q}$ has eigenvalues\citep{laubMatrixAnalysisScientists2005}
\begin{equation}
    \lambda_{pq} = \mu_p - \mu_q, \quad p,q = 1, \ldots, N.
\end{equation}
We see that $\mbf{Q}$ has at least $N$ eigenvalues equal to zero and so
$\mbf{Q}$ is always singular. 
Now assume that $\mbf{K}$ is diagonalizable as
\begin{align} \label{eq:kappa_diag}
    \mbf{K} &= \mbf{R} \bm{\mu} \mbf{L}^{\dagger}, \quad 
    \mbf{L}^{\dagger} \mbf{R} = \mbf{R} \mbf{L}^{\dagger} = \mbf{1}
\end{align}
where $\bm{\mu}$ is a diagonal matrix holding the eigenvalues $\mu_p$. 
In the notation
of Eq.~\eqref{eq:kappa_diag}, the rows of $\mbf{L}^{\dagger}$ are the left eigenvectors 
of $\mbf{K}$ while the columns of $\mbf{V}$ are the right eigenvectors of $\mbf{K}$. 
In addition, let
\begin{equation}
    \bm{\Lambda} = \bm{\mu} \otimes \mbf{1} - \mbf{1} \otimes \bm{\mu}
\end{equation}
be the diagonal matrix holding the eigenvalues $\lambda_{pq} = \mu_p - \mu_q$ of $\mbf{Q}$.
It follows directly that
\begin{align}
    (\mbf{L}^{*} \otimes \mbf{R})
    \bm{\Lambda}
    (\mbf{R}^{\trans} \otimes \mbf{L}^{\dagger})
    &= (\mbf{L}^{*} \otimes \mbf{R})
    (\bm{\mu} \otimes \mbf{1} - \mbf{1} \otimes \bm{\mu})
    (\mbf{R}^{\trans} \otimes \mbf{L}^{\dagger}) \nn
    &= (\mbf{L}^{*} \bm{\mu} \mbf{R}^{\trans}) \otimes (\mbf{R} \mbf{L}^{\dagger})
    -  (\mbf{L}^{*} \mbf{R}^{\trans}) \otimes (\mbf{R} \bm{\mu} \mbf{L}^{\dagger}) \nn
    &= \mbf{K}^{\trans} \otimes \mbf{1} - \mbf{1} \otimes \mbf{K} \nn
    &= \mbf{Q}.
\end{align}
We see that $\mbf{Q}$ can be diagonalized in a simple way provided $\mbf{K}$ is 
diagonalizable. The matrices
\begin{subequations}
    \begin{align}
        \bm{\mathfrak{R}} &= \mbf{L}^{*} \otimes \mbf{R} \\
        \bm{\mathfrak{L}}^{\dagger} &= \mbf{R}^{\trans} \otimes \mbf{L}^{\dagger}
    \end{align}
\end{subequations}
contain the right and left eigenvectors of $\mbf{Q}$, respectively, and satisfy
$\bm{\mathfrak{L}}^{\dagger} \bm{\mathfrak{R}} = \bm{\mathfrak{R}}  \bm{\mathfrak{L}}^{\dagger} = \mbf{1}$
due to Eq.~\eqref{eq:kappa_diag}. Following Eq.~\eqref{eq:main_d_using_diag}, we are now ready to compute
\begin{subequations}
    \begin{alignat}{2}
        \mbf{D} &= \bm{\mathfrak{R}} \varphi(\bm{\Lambda}) \bm{\mathfrak{L}}^{\dagger}
        &&= \bm{\mathfrak{R}} \mbf{\Phi} \bm{\mathfrak{L}}^{\dagger}, \\
        \mbf{D}^{\trans} &= \bm{\mathfrak{L}}^{*} \varphi(\bm{\Lambda}) \bm{\mathfrak{R}}^{\trans}
        &&= \bm{\mathfrak{L}}^{*} \mbf{\Phi} \bm{\mathfrak{R}}^{\trans}
    \end{alignat}
\end{subequations}
where the diagonal matrix $\mbf{\Phi}$ has diagonal elements
\begin{align}
    \varphi_{pq} = \varphi(\lambda_{pq}) = \varphi(\mu_p - \mu_q).
\end{align}
If $\mbf{D}$ and thus $\mbf{D}^{\trans}$ is invertible, we write
\begin{align}
    \mbf{D}^{-\trans} 
    = \bm{\mathfrak{L}}^{*} \mbf{\Phi}^{-1} \bm{\mathfrak{R}}^{\trans}
\end{align}
and proceed to compute the matrix-vector product
\begin{align}
    \dot{\bm{\kappa}} 
    &= -i \mbf{D}^{-\trans} \mbf{g} \nn
    &= -i \bm{\mathfrak{L}}^{*} \mbf{\Phi}^{-1} \bm{\mathfrak{R}}^{\trans} \mbf{g} \nn
    &= -i (\mbf{R} \otimes \mbf{L}^{*}) \mbf{\Phi}^{-1} (\mbf{L}^{\dagger} \otimes \mbf{R}^{\trans}) \mbf{g}.
\end{align}
This exact type of product was encountered in the appendix of Ref. \citenum{hojlundBivariationalTimedependentWave2022} and can be simplified by
using that
\begin{equation} \label{eq:ABC}
    (\mbf{A} \otimes \mbf{B}) \mrm{vec}(\mbf{C}) = \mrm{vec}(\mbf{A} \mbf{C} \mbf{B}^{\trans})
\end{equation}
where $\mrm{vec}(\cdot)$ denotes the row-major vectorization mapping. Rather than repeating the
steps, we simply state the result:
\begin{align}
    \dot{\mbf{K}}
    = - i \mbf{R} \big( \mbf{\Omega} \circ \big(\mbf{L}^{\dagger} \mbf{G} \mbf{R} \big) \big) \mbf{L}^{\dagger}.
\end{align}
Here, $\dot{\mbf{K}}$ and $\mbf{G}$ are the matrices holding the elements of the vectors $\dot{\bm{\kappa}}$
and $\mbf{g}$, respectively, while the symbol $\circ$ denotes the Hadamard (or element-wise) product.
We have, in addition, introduced the $N \times N$ matrix $\mbf{\Omega}$ that holds the $N^2$ diagonal elements of the $N^2 \times N^2$ diagonal matrix $\mbf{\Phi}^{-1}$, i.e.
\begin{align}
    \omega_{pq} = 1 / \varphi(\lambda_{pq}) = 1 / \varphi(\mu_p - \mu_q).
\end{align}
This result holds, as mentioned, when $\mbf{D}^{\trans}$ is invertible. According to the discussion in Sec.~\ref{ssec:similarity_transformed_derivative},
this is the case exactly when
\begin{align}
    \mu_p - \mu_q \neq 2\pi i k
\end{align}
where $k$ is a non-zero integer. 
\newpage

\end{document}


\title{General exponential basis set parametrization: Application to time-dependent bivariational wave functions: Supplementary material}
\author{Mads Greisen Højlund}
\email{madsgh@chem.au.dk}
\affiliation{\au}
\author{Alberto Zoccante}
\email{alberto.zoccante@uniupo.it}
\affiliation{\upo}
\author{Ove Christiansen}
\email{ove@chem.au.dk}
\affiliation{\au}

\hypersetup{pdftitle={General exponential basis set parametrization: Application to time-dependent bivariational wave functions: Supplementary material}}
\hypersetup{pdfauthor={M.~G.~Højlund, et al.}}
\hypersetup{bookmarksopen=true}

\date{March 30, 2023}
\maketitle


\renewcommand{\thefigure}{S\arabic{figure}}

\section{Additional figures: water} \label{sec:water}

\begin{figure}[H]
    \centering
    \includegraphics[width=1.1\columnwidth, angle=90]{{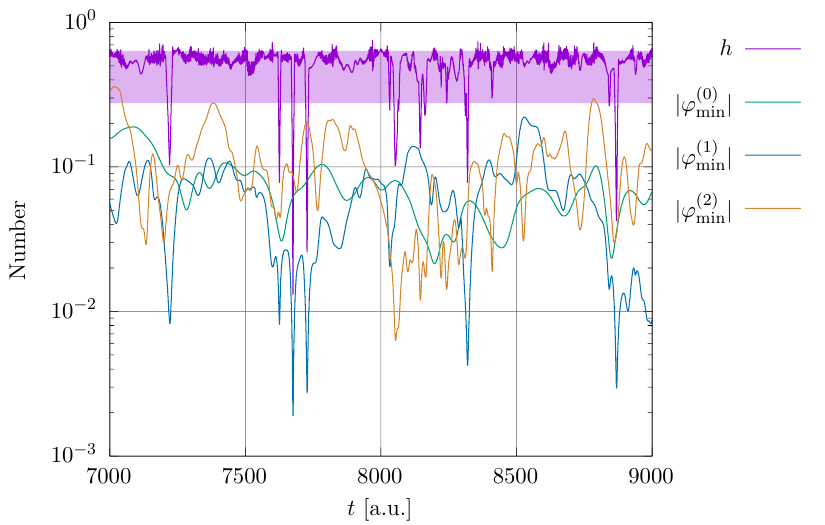}}
    \caption{Integrator step size ($h$) and $|\varphi^m_\mathrm{min}|$ for the intramolecular vibrational energy redistribution (IVR) of water described
    at the TDMVCC$[2]$ level using double exponentially parametrized modals ($\tau = 0$, i.e. no resets).
    The shaded area covers the mean step size plus/minus its standard deviation (computed within the shown time interval).
    Only part of the full time interval is shown for greater visibility.}
    \label{fig:water_tdmvcc2_double_exp_stepsize}
\end{figure}

\begin{figure}[H]
    \centering
    \includegraphics[width=1.1\columnwidth, angle=90]{{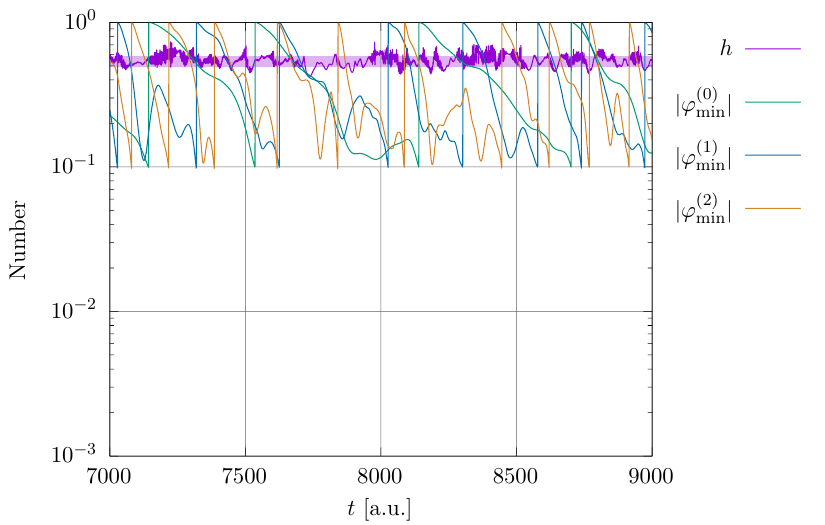}}
    \caption{Integrator step size ($h$) and $|\varphi^m_\mathrm{min}|$ for the intramolecular vibrational energy redistribution (IVR) of water described
    at the TDMVCC$[2]$ level using double exponentially parametrized modals ($\tau = 0.1$).
    The shaded area covers the mean step size plus/minus its standard deviation (computed within the shown time interval).
    Only part of the full time interval is shown for greater visibility.}
    \label{fig:water_tdmvcc2_double_exp_trans1.0e-1_stepsize}
\end{figure}

\section{Additional figures: 5D \textit{trans}-bithiophene} \label{sec:tbithio}

\begin{figure}[H]
    \centering
    \includegraphics[width=1.1\columnwidth, angle=90]{{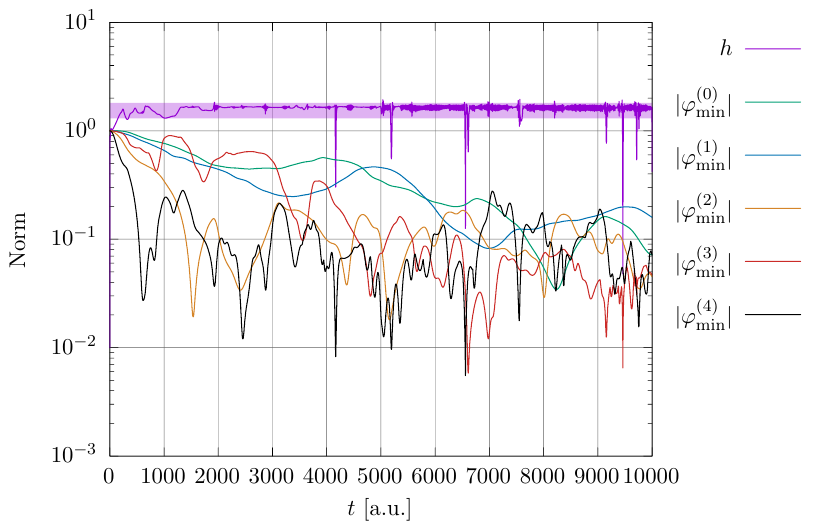}}
    \caption{Integrator step size ($h$) and $|\varphi^m_\mathrm{min}|$ for the 5D \textit{trans}-bithiophene model described
    at the TDMVCC$[2]$ level using single exponentially parametrized modals ($\tau = 0$, i.e. no resets).
    The shaded area covers the mean step size plus/minus its standard deviation.}
    \label{fig:tbithio_tdmvcc2_exp_stepsize}
\end{figure}

\begin{figure}[H]
    \centering
    \includegraphics[width=1.1\columnwidth, angle=90]{{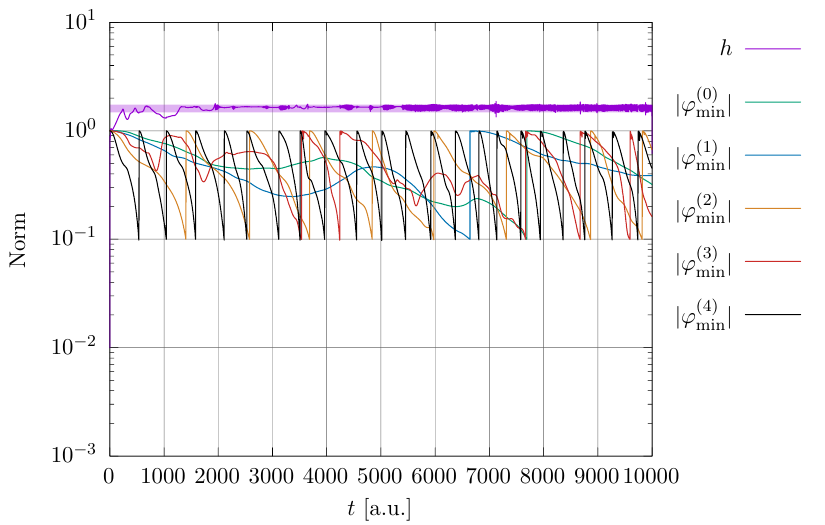}}
    \caption{Integrator step size ($h$) and $|\varphi^m_\mathrm{min}|$ for the 5D \textit{trans}-bithiophene model described
    at the TDMVCC$[2]$ level using single exponentially parametrized modals ($\tau = 0.1$).
    The shaded area covers the mean step size plus/minus its standard deviation.}
    \label{fig:tbithio_tdmvcc2_exp_trans1.0e-1_stepsize}
\end{figure}